\documentclass[%
 reprint,
superscriptaddress,
 amsmath,amssymb,
 aps,
pra, 
longbibliography,
floatfix ]{revtex4-1}

\usepackage{natbib} 
\usepackage{amsmath} 
\usepackage{graphicx} 
\usepackage[usenames,dvipsnames]{color}  
\usepackage{longtable}
\usepackage{multirow}
\usepackage{amsmath}
\usepackage{epstopdf}

\usepackage[english]{babel}

\makeatletter
\def\bbl@set@language#1{%
  \edef\languagename{%
    \ifnum\escapechar=\expandafter`\string#1\@empty
    \else\string#1\@empty\fi}%
  \@ifundefined{babel@language@alias@\languagename}{}{%
    \edef\languagename{\@nameuse{babel@language@alias@\languagename}}%
  }%
  \select@language{\languagename}%
  \expandafter\ifx\csname date\languagename\endcsname\relax\else
    \if@filesw
      \protected@write\@auxout{}{\string\select@language{\languagename}}%
      \bbl@for\bbl@tempa\BabelContentsFiles{%
        \addtocontents{\bbl@tempa}{\xstring\select@language{\languagename}}}%
      \bbl@usehooks{write}{}%
    \fi
  \fi}
\newcommand{\DeclareLanguageAlias}[2]{%
  \global\@namedef{babel@language@alias@#1}{#2}%
} \makeatother

\DeclareLanguageAlias{en}{english}

\begin{document}

\title{Multi-Dimensional Item Response Theory and the Force Concept Inventory}

\author{John Stewart}
\author{Cabot Zabriskie}%
\author{Seth DeVore}%
\author{Gay Stewart}
\affiliation{%
Department of Physics and Astronomy, West Virginia University,
Morgantown WV, 26506
}%
\email{jcstewart1@mail.wvu.edu}
\date{\today}

\begin{abstract}

Research on the test structure of the Force Concept Inventory
(FCI) has largely been performed with exploratory methods such as
factor analysis and cluster analysis. Multi-Dimensional Item
Response Theory (MIRT) provides an alternative to traditional
Exploratory Factor Analysis which allows statistical testing to
identify the optimal number of factors. Application of MIRT to a
sample of $N=4,716$ FCI post-tests identified a 9-factor solution
as optimal. Additional analysis showed that a substantial part of
the identified factor structure resulted from the practice of
using problem blocks and from pairs of similar questions. Applying
MIRT to a reduced set of FCI items removing blocked items and
repeated items produced a 6-factor solution; however, the factors
had little relation the general structure of Newtonian mechanics.
A theoretical model of the FCI was constructed
from expert solutions and fit to the FCI by constraining the MIRT
parameter matrix to the theoretical model. Variations on the
theoretical model were then explored to identify an optimal model.
The optimal model supported the differentiation of Newton's 1st
and 2nd law; of one-dimensional and three-dimensional kinematics;
and of the principle of the addition of forces from Newton's 2nd
law. The model suggested by the authors of the FCI was also fit;
the optimal MIRT model was statistically superior.

\end{abstract}

\maketitle

\section{Introduction}

The Force Concept Inventory (FCI) was introduced 25 years ago and
has become one of the most used and most studied instruments in
Physics Education Research (PER) \cite{fci}. Measurements using
the instrument have been important in the recognition that
traditional instruction was not sufficient for students to develop
a conceptual understanding of Newton's laws \cite{hake1998}. Its
success was followed by the development of numerous other
conceptual instruments some of which found wide-spread use
including the Force and Motion Conceptual Evaluation \cite{thornton1998},
the Conceptual Survey of Electricity and Magnetism
\cite{maloney2001}, and the Brief Electricity and Magnetism
Assessment \cite{ding2006}. These four instruments have in turn
been used to help understand the effect of pedagogical
innovations, the challenges of learning physics, and issues of
inclusion in physics. The impact of these instruments has been
immense; they have been used in a substantial subset of the
studies done in PER. For a broad overview of PER including the
role of conceptual inventories in PER, see Docktor and Mestre's
recent synthesis \cite{docktor2014synthesis}.

A substantial number of studies have attempted to understand the
overall structure of the FCI. These have included purely
exploratory or descriptive methods such as factor analysis
\cite{huffman_what_1995,scott2012exploratory,semak2017}, module
analysis \cite{brewe_using_2016}, cluster analysis
\cite{springuel2007applying,stewart2012}, item response theory
\cite{wang_analyzing_2010,planinic2010,popp2011,traxler2017} and
item response curves \cite{morris2006testing,morris2012item}. The
structure of student reasoning on the FCI has also been
investigated by methods such as model analysis that require the
input of a partial model of the concepts measured by the FCI
\cite{bao2006}. Model analysis was later shown to be exact only in
certain limiting cases \cite{stewart2012}. For a summary of these
exploratory and non-exploratory methods, see the review by Ding and
Beichner \cite{ding2009}.

The reliability and validity of the FCI have also been tested. The
internal consistency of the FCI measured by Cronbach's alpha is
quite strong \cite{lasry2011puzzling,traxler2017}. The instrument
has also demonstrated good test-retest reliability
\cite{henderson2002common}. While some validity issues have been
identified \cite{traxler2017}, these are minor compared to those
reported for some other instruments \cite{jorion_analytic_2015}.

The current study explored the factor structure of the FCI using
Multi-Dimensional Item Response Theory (MIRT). This method has
previously been applied to the FCI \cite{scott2015} by Scott and
Schumayer. MIRT, described in detail in Sec. \ref{sec:methods},
provides statistical criteria for determining the optimal number
of factors unlike traditional Exploratory Factor Analysis (EFA). The current study
applies MIRT to the FCI using a larger dataset than the previous study collected under conditions
where correct answering is more strongly incentivized, thus allowing a finer resolution
of the details of student thinking.
It also allows models to be constrained to eliminate factor
loadings that should not theoretically occur. As such, it allows a
more detailed exploration of the structure of an instrument than
traditional EFA.

\subsection{Factor Analysis and the FCI}

The authors' of the FCI provided a detailed description of the
physical concepts each item in the original instrument was
designed to measure \cite{fci}. Soon after its publication,
attempts to extract the suggested structure with EFA were
unsuccessful leading to debate about what the instrument actually
measured
\cite{huffman_what_1995,hestenes1995interpreting,heller1995interpreting}.
Huffman and Heller reported that, for a sample of 145 high school
students, principle component analysis identified two factors:
Newton's 3rd law and Kinds of Forces. For 750 university students,
only one factor was identified: Kinds of Forces. This study
selected the number of factors by requiring that each new factor
explain at least 5-10\% additional variance. The difference in the
number of factors identified between the Huffman and Heller study
and other studies of the FCI may have resulted from the use of
different criteria to identify the optimal number of factors. Methods to identify
the optimal number of factors are discussed in Sec. \ref{sec:analysis}.

Scott, Schumayer, and Gray applied EFA to the FCI post-test scores
of a sample of 2,150 students in a college algebra-based physics
course \cite{scott2012exploratory}. The FCI was delivered
electronically and students were given no special incentive for
completion. They found a single factor explained a substantial
portion of the variance, but concluded a five-factor model was
optimal. Parallel analysis was used to select the optimal number
of factors. The ``knee'' of their Scree plot suggested that two or
three factors could also be considered optimal. In examining the
loadings on the single factor, they discuss the possibility of
very difficult items not being strongly correlated with the
single-factor solution. The variance explained by the addition of
each new factor is not reported, and therefore, the number of
factors selected cannot be compared with Huffman and Heller's
solution \cite{huffman_what_1995}.

Semak {\it et al.} explored the evolution of the structure of
student thinking on the FCI using factor analysis \cite{semak2017}
for 427 algebra- and calculus-based introductory physics students.
They found the optimal solution had 5 factors on the pretest and 6
factors on the post-test. Parallel analysis was used to select the
optimal number of factors; however, examination of the Scree plots
from their study suggests one could have also identified one or
two factors as optimal for both the pretest and post-test. This
would have provided support for Huffman and Heller's model. We
provide a comparison of the four reported factor structures in
Sec. \ref{sec:discussion}.

Factor analysis has also been used to investigate other sets of
physics problems. Ramlo \cite{ramlo2008validity} calculated the
factor structure of the FMCE \cite{thornton1998} finding 3 factors
for the pretest; however, these factors contained a mixture of
concepts and Ramlo concluded the pretest factor structure was
undefined. Three factors were also found for the post-test with
items covering similar conceptual topics largely loading onto the
same factor. Ramlo used a Scree plot to identify an eigenvalue
cutoff of 2.5 to determine the optimal number of factors.

\subsection{Item Response Theory and the FCI}

Item Response Theory (IRT) contains a broad set of statistical
models which calculate the probability of a student with some
overall proficiency or ability to answer individual items on a
test correctly. Many different IRT models have been used to
investigate the FCI including the Rasch model, the 2PL model, the
3PL model, and MIRT. These models are reviewed in Sec.
\ref{sec:methods}.

Many studies have investigated the FCI with IRT using a single
ability parameter (unidimensional IRT). Wang and Bao employed the
3PL IRT model to investigate the FCI pretest for 2,802 college
students taking calculus-based physics \cite{wang_analyzing_2010}.
They reported excellent model fit with all items showing
reasonable difficulty parameters and no items with negative
discrimination parameters. The 3PL model adds a parameter to the
2PL model to account for random guessing. The majority of the
guessing parameters were less than the $20\%$ random guessing
would produce. The use of the 3PL model for distractor-driven
instruments has been questioned \cite{morris2012item}.

Planinic, Ivanjek, and Susac performed a Rasch analysis of 1,676
Croatian high school students who had completed an algebra-based
physics class \cite{planinic2010}. The Rasch model difficulty
parameters were largely in agreement with the overall item
average. This study is difficult to generalize because the overall
score on the instrument (27.7\%) was so low and the measurement
was performed two and one-half years after instruction.

Osborn Popp, Meltzer, and Megowan-Romanowicz also used Rasch Model
IRT for a sample of 4,775 high school students to investigate item
fairness; all students had been taught using Modeling Instruction
\cite{popp2011}. IRT using the Rasch model was used to determine
if items within the FCI were of equal difficulty for men and
women. They found that a number of items were significantly easier
for male students and some for female students.

Traxler {\it et al.} \cite{traxler2017} also investigated item
fairness in the FCI with IRT using the 2PL model. They found that
eight items were substantially biased toward men and and two
toward women; they proposed a reduced 19-item instrument to
eliminate all biased and poorly functioning items.

Han {\it et al.}  used the 3PL IRT model as part of the process of
evaluating the equivalence of two shorter versions of the FCI
\cite{han2015}. Traxler {\it et al.} \cite{traxler2017} cautions
that the gender unfair items were not evenly distributed between
the shortened tests, and therefore, the two shorter tests might
have different performance for men and women.

Scott and Schumayer \cite{scott2015} attempted to replicate the
work of Scott, Schumayer, and Gray \cite{scott2012exploratory} on
a related dataset using MIRT. They confirmed the 5-factor
solution. Comparing the factor models of the two studies showed
very good, but not perfect, agreement suggesting MIRT and EFA are
complementary techniques. To select the optimal number of factors,
AIC and BIC (described in Sec. \ref{sec:methods}) were minimized.

IRT has also been used to explore other sets of physics problems.
Lee {\it et al.} \cite{lee2008measuring} used 2PL IRT to examine
how
 the skill of physics students using an
online homework system changed between their first and second
attempts at a problem. Whether feedback was given on the first
attempt and the type of feedback strongly influenced the change in
student skill (IRT ability) between the first and second attempt.

Morris {\it et al.} \cite{morris2006testing} introduced an
alternative to IRT (bearing a very similar name), Item Response
Curves (IRC), which was  used to analyze the FCI. IRC analysis
simplifies IRT analysis by using the overall test score as a
surrogate for student ability, greatly reducing computational
demands and allowing intuitive exploration of the effect of
distractors. Using a sample of over 4,500 students drawn from
multiple institutions, a later study by Morris {\it et al.}
\cite{morris2012item} compared IRC analysis to the IRT analysis of
Wang and Bao \cite{wang_analyzing_2010} and found excellent
correlation between the difficulty parameters of the models.

\subsection{The Structure of Knowledge}

Most explorations of the structure of the FCI have focused on
determining a general structure which represents the entire
instrument in terms a small number of factors/clusters. This
reductionism is at odds with a large body of research suggesting
students' knowledge of physics is complex and that students
(novices) do not possess the strongly integrated view of physics
of expert practitioners. Expert and novices categorize problems
differently; novices by surface features and experts by deeper
conceptual divisions \cite{gliner1,gliner2}. One commonly accepted
difference in the knowledge structure of experts is the
hierarchical nature of the structure, with the most fundamental
principles at the top and less fundamental concepts branching out
from there \cite{eylon1984effects, chi1981categorization,
schoenfeld1982problem}. This more deliberate structuring of
knowledge allows experts to engage more efficiently in chunking of
knowledge \cite{reif1982knowledge, beatty2002probing,
miller1956magical} for more expedient application of the correct
physics principles when engaging in problem solving.

Conversely, novices lack this deliberate knowledge structure
leading to less deliberate methods of problem solving. This review
will follow the categorization of expert/novice research presented
in Docktor and Mestre's extensive synthesis of PER
\cite{docktor2014synthesis}. One view regarding some of the novel
ways that novices approach problems differently from experts is
the ``misconceptions'' view. This view argues that students,
through their life experiences, have developed theories regarding
how the world works and that using these, often incorrect, theories
leads to some of the common difficulties in physics problem
solving \cite{clement1982students, mcdermott1984research,
posner1982accommodation}. Research into these misconceptions has
shown that they are very difficult to overcome due in part to the
time students have spent believing them to be true
\cite{etkina2005cognitive, national2000people}. Another method of
explaining the differences is the ``ontological categories'' view,
which posits that students miscategorize their knowledge, storing
it in incorrect broad categories (i.e. thinking of force as a
thing that can be used up) \cite{chi1993ontological,
chi1994things, slotta1995assessing}. Another popular theoretical
framework is the ``knowledge in pieces'' view
\cite{disessa1993toward, disessa1998changes,
hammer1996misconceptions} wherein student understanding consists
of a number of granular facts that are activated, either
individually or in small groups, to synthesize a solution.
Regardless of the theoretical framework used to describe it,
novice knowledge and the associated problem solving techniques
have been shown to be highly sensitive to the context of the problem
and how it relates to problems they have seen in the past
\cite{gillespie2004coherence, dufresne2002marking,
steinberg1997performance}. As such, the knowledge state of
students may be better described by models of a granular knowledge
structure instead of the integrated models implied by factor
analysis or cluster analysis.

The current work will produce a fine-grained model of the
information needed to solve FCI problems. This model is very
similar to models produced by a paradigm of cognitive research
into complex problem solving pioneered by Simon and Newell
\cite{simonnewell}. This paradigm and its history, which
dominated research into problem solving for over 30 years, were
summarized by Ohlsson \cite{ohlsson2012problems}. The paradigm
constructed computational models that replicated the problem
solving sequence of human solvers; the sequence of the human
solver was identified by coding extensive think-aloud transcripts.
This method was applied to examine expert/novice differences in
problem solving in kinematics and dynamics, as well as other
fields \cite{larkin1980expert,larkin1980models}. Reif and Heller
offered a related detailed model of problem solving in mechanics
\cite{reif1982knowledge}; this model did not meet the test of
being computationally functional, but was meant to be a complete
model that could serve as a prescription of expert behavior. The
model we will propose for the FCI shares many features with the
computational models of Larkin {\it et al.}
\cite{larkin1980models} and the model of Reif and Heller
\cite{reif1982knowledge}. The work on complex problem solving
summarized focussed primarily on quantitative solutions; however,
the framework presented by Reif and Heller acknowledged the role
of qualitative decisions in the solution process and suggested
extensions to model qualitative reasoning.

\subsection{Research Questions}

This study seeks to answer the following research questions.
\begin{enumerate}
\item[RQ1:] What factor structure is extracted for the FCI by
MIRT? Is this structure consistent with the results of other
factor analysis? \item[RQ2:] Can parts of this factor structure be
explained by factors other than the structure of student knowledge
of Newtonian mechanics? \item[RQ3:] If blocked items and repeated
reasoning groups are removed, is the resulting factor structure
consistent with Newtonian mechanics? \item[RQ4:] Can theoretically
constrained MIRT produce a model of the physical constructs
measured by the FCI? If so, what is the optimal model of the FCI
for this student population? \item[RQ5:] Does the structure
proposed by the FCI's authors provide a superior description of
the instrument to the optimal model identified by MIRT?
\end{enumerate}

This work leaves two important areas of analysis for future
research: the role of misconceptions and bias. The FCI was
constructed so that the distractors represented common
misconceptions. In the analysis in this paper, only the
correctness of the responses was analyzed. MIRT could be
extended to include factors representing common misconceptions to
determine how the models presented in this work would be modified.

There is a substantial body of research indicating that some
problems within the FCI are unfair to female students with a few
unfair to male students. These problems have often factored
together in previous analysis
\cite{scott2012exploratory,semak2017} leading to the possibility
that some factors are identified because of biases in the
problems. Many biased problems were removed in the analysis in
this study to remove spurious correlations; however, future
research should investigate whether the factor structure
identified is independent of gender. While this study will not focus on
gender fairness, the reduced fair 19-item FCI proposed by Traxler {\it et al.} \cite{traxler2017}
will be examined using the optimal theoretical FCI model identified by MIRT. For a review of  research
into FCI item bias see Traxler {\it et al.} \cite{traxler2017}.
For a review of the issue of gender disparities in conceptual
inventories see Madsen, McKagan, and Sayre
\cite{madsen_gender_2013} or Henderson {\it et al.}
\cite{henderson2017}. For a general review of gender in physics
see Traxler {\it et al.} \cite{traxler_enriching_2016}.

\section{Methods}
\label{sec:methods}

\subsection{Force Concept Inventory}
The FCI is a 30-item multiple-choice instrument that includes
conceptual questions about Newton's laws, kinematics, and forces
\cite{fci}. Each item has five possible responses. The incorrect
responses were developed to include common misconceptions. The FCI
contains some individual items and some items that are grouped
into blocks which share a common stem. The FCI was revised after
its introduction; this work will use the revised FCI published
with Mazur's {\it Peer Instruction} \cite{mazur1996peer} and
available at PhysPort \cite{physport}.

\subsection{Sample}

The sample of FCI post-test results was collected from a large,
southern land-grant university with an enrollment of approximately
25,000 students. This university held a Carnegie Classification of
``Highest Research Activity'' for the period studied. The sample
is comprised of 4,716 complete post-test responses collected from
the spring 2002 semester to fall 2012 semester (23.1\% women). The
demographics of the university in 2012 were 79\% White, 5\%
African American, 6\% Hispanic, and 3\% or less of other groups.
The 25th to 75th percentile range of the general student
population's ACT scores was 23-29 \cite{usnews}. This sample was
also used in the analysis of Traxler {\it et al.}
\cite{traxler2017}.

The sample was collected in the introductory calculus-based
mechanics course serving future physical scientists and engineers.
Students in the course were required to attend two 50-minute
lectures and two two-hour laboratories each week. The lectures
were largely traditional with attendance monitored by a quiz given
at the beginning and end of each session. The lab sessions featured
a mixture of activities including teaching assistant (TA) led
interactive demonstrations, small group problem solving,
inquiry-based hands-on activities, and traditional experiments.
The class had been revised previous to the beginning of data
collection and was presented with few changes over the period
studied. The class was managed by the same lead instructor for the
period studied; this instructor taught 75\% of the lecture
sections and oversaw the instruction of the remaining sections.

\subsection{Item Response Theory}

Many IRT probability models have been constructed to model student
responses to different test structures and testing situations
\cite{crcbook}. One of the most intuitive and widely used is the
two-parameter logistic (2PL) model. The 2PL model uses
unidimensional IRT which explicitly models the effect of the
single latent trait of ability, $\theta_i$, on the probability of
a student, $i$, successfully answering a question. The 2PL model
assumes that each item $j$ has a difficulty $b_j$ and
discrimination $a_j$. The probability, $\pi_{ij}$, that student
$i$ will successfully answer item $j$ is given by the logistic
function:
\begin{equation}
\label{eqn:2pl}
\pi_{ij}=\frac{\exp[a_j(\theta_i-b_j)]}{1+\exp[a_j(\theta_i-b_j)]}.
\end{equation}
The 2PL model can be expanded to the 3PL model by including an
additional parameter for each item to model random guessing. The
3PL model has been used in some studies of the FCI. A
simplification of the 2PL model, called the Rasch model, has also
been used to study the FCI; the Rasch model sets the
discrimination of each item to one, $a_j=1$.

The assumption of unidimensional IRT, that a single ability
parameter captures the students' facility with the test material,
may be correct for some instruments but it seems unlikely for the
FCI, which measures a number of different facets of Newton's laws
and kinematics. Multi-Dimensional IRT (MIRT) extends
unidimensional IRT by estimating multiple ability traits for each
student. Mathematically, the student ability $\theta_i$ which is a
scalar in unidimensional IRT becomes a vector,
$\boldsymbol{\theta_i}$, in MIRT. If $k$ ability traits are
estimated for each item, each trait is associated with its own
item discrimination, $a_{jk}$, making the discrimination a vector,
$\boldsymbol{a_j}$.

Multiple MIRT models exist; the most common MIRT model is call the
\textit{ compensatory model} where the probability of a particular
response is determined by a linear combination of
$\boldsymbol{\theta_i}$ components where large components of
$\boldsymbol{\theta_i}$ will compensate for the smaller components
of $\boldsymbol{\theta_i}$. This model is shown in Eqn.
\ref{eqn:2plCMIRT}
\begin{equation}
\label{eqn:2plCMIRT}
\pi_{ij}=\frac{\exp[\boldsymbol{a_j}\cdot\boldsymbol{\theta_i}+d_j]}{1+\exp[\boldsymbol{a_j}\cdot\boldsymbol{\theta_i}+d_j]},
\end{equation}
where $d_j$ would be the product $-a_jb_j$ in the 2PL model and the
product $\boldsymbol{a_j}\cdot\boldsymbol{\theta_i}$ is a dot
product of two vectors. The parameter $d_j$ is related to the
difficulty of the item. While this MIRT model estimates multiple
discrimination parameters for each item, it estimates only one
$d_j$ parameter. This is not optimal; it would be beneficial to
know the difficulty of the item by individual trait.
Non-compensatory MIRT would extract the difficulty of each item;
however, this doubles the number of parameters estimated. We
attempted to apply non-compensatory MIRT to the FCI but the models
did not converge.

\subsection{Model Fit Statistics}

Unlike traditional factor analysis which identifies factors as
eigenvectors of the correlation matrix, IRT introduces a
statistical model which is then fit to the observations. The model
is used to calculate the likelihood function, $L$, which
represents the probability the observation occurred given the
probability model. Maximum likelihood estimation techniques are
used to search the parameter space to select a set of parameters
which maximize $L$, the set of parameters which make the observed
results most likely. This form of estimation can be used for a
wide set of models and a number of general model fit statistics
have been developed. We will report the Akaike Information
Criterion (AIC), AIC$=2k-2\ln(L)$, and the Bayesian Information
Criterion (BIC), BIC$=k\ln(n)-2\ln(L)$, where $n$ is the sample
size and $k$ the number of parameters estimated. The optimal model
minimizes both quantities; BIC penalizes the addition of
parameters more strongly than AIC. Because AIC and BIC are
calculated from the logarithm of $L$, a change of 2.3 units in
either represents a change of $e^{2.3}=10$ in the likelihood of
the model, an order of magnitude.

A substantial number of additional fit statistics have been
developed for maximum likelihood models. We will report the Root
Mean Square Error of Approximation (RMSEA), the Comparative Fit
Index (CFI), and the Tucker-Lewis Index (TLI). Hu and Bentler
suggest using multiple indices to evaluate model fit
\cite{hu1999}. Acceptable model fit is characterize by
RMSEA$<0.05$ and CFI$>0.96$ or TLI$>0.96$.

\subsection{Additional Analysis}
\label{sec:analysis}
While MIRT allows statistical selection of the optimal number of
factors, traditional EFA uses a number of non-statistical
criteria. Factor selection often begins by an examination of the
``Scree'' plot which plots the eigenvalue of the
correlation/covariance matrix corresponding to the factor against
the factor number; the eigenvalue is related to the variance
explained by the factor. An example of a Scree plot is shown in
the Supplemental Material \cite{supp}. One, then, identifies the
``knee'' of the Scree plot, the point of greatest curvature. The
number of factors corresponding to the knee is the optimal number
of factors. For factor numbers greater than the location of the
knee, each additional factor explains substantially less variance
than the factors already extracted.

Many additional methods have been developed and often yield
contradictory results. The sum of the eigenvalues of
the correlation/covariance matrix is equal to the trace of the
matrix; therefore, an eigenvalue that is less than the mean
correlation/covariance represents a factor that explains less
variance than an individual item. The optimal number of factors
can then be extracted as the last factor with eigenvalue above the
mean. Parallel analysis computes the eigenvalues of a random
correlation matrix; the optimal factor number is the last factor
with eigenvalue greater than the parallel analysis eigenvalue.

Partial correlation matrices will be reported. The partial
correlation matrix for the dichotomous scores on individual FCI
items was calculated by regressing the total FCI score on the
individual item score using a general linear model. The
correlations of the residuals of these regressions form the
partial correlation matrix.

The mean and standard deviation of MIRT parameters,
$\boldsymbol{a_j}$ and $d_j$, were calculated by bootstrapping.
Bootstrapping is a statistical technique that allows the
calculation of the average, standard deviation, and confidence
interval without assuming a statistical model. This is done by
forming sub-samples of the data with replacement and recalculating
the desired parameter for each sub-sample. For this work, 200
sub-samples were used; this calculation required one week of
computational time on a modern personal computer.

All statistical analysis were carried out in the ``R'' software
package \cite{rsoftware}. MIRT was performed with the ``mirt''
package \cite{mirt}. This work used correlation analysis to
investigate the origin of the factor structures extracted. The
correlation matrix will be presented in a visualization rendered
by the ``qgraph'' package \cite{qgraph}. Partial correlation
matrices will be constructed by using the ``glm'' function to
regress the total FCI score on the dichotomous scores of the
individual items. Factor analysis was
carried out with the ``factanal'' function in the ``stats''
package. The ``nFactors'' package was used to generate the Scree
plot and to perform parallel analysis. Bootstrapping was performed
by the ``boot'' package \cite{boot1,boot2}.

\subsection{Supplemental Materials}
See Supplemental Material \cite{supp}
for traditional factor analysis including the Scree plot, 3- and 5-factor MIRT  models, and the constrained MIRT model without the
factor loading on all items \cite{supp}.

\section{Results}
The FCI was first examined with MIRT without employing any
theoretical model, thus performing an EFA. Correlation analysis was
then used to understand the resulting factor structure. Expert
solutions of the FCI were then used to construct a theoretical
model of the instrument. This model further explained the
correlation structure observed. MIRT was then used to explore how
the theoretical model mapped onto student responses to the FCI.
Finally, the model proposed by the FCI authors was fit and
compared to the optimal model in this work.

\subsection{Exploratory Factor Analysis}

MIRT was used to perform a factor analysis on the FCI. Models with
progressively more factors were fit and compared using ANOVA. A
9-factor model improved model fit over an 8-factor model
[$\chi^2(22)=53.44$, $p<0.001$] and explained $56\%$ of the
variance in the item scores. The last factor added explained
$3.6\%$ additional variance. The 10-factor model did not
significantly improve model fit. The 9-factor model (varimax
rotation) is shown in Table \ref{tab:mirt-full}. Factors are reported as columns and labelled ``FC.'' The table also
identifies the FCI problem blocks. The table reports, $d$; $d$ is related to the
overall difficulty of the item (Eqn. \ref{eqn:2plCMIRT}); easier
items have larger $d$.

\begin{table}[!tbp]
\caption{Factor Loadings for Exploratory Factor Analysis with
Multi-dimensional IRT (varimax rotation). \label{tab:mirt-full}}
\begin{center}
\begin{tabular}{llllllllllr}
\hline\hline
\multicolumn{1}{l}{\#}&\multicolumn{1}{c}{FC1}&\multicolumn{1}{c}{FC2}&\multicolumn{1}{c}{FC3}&\multicolumn{1}{c}{FC4}&\multicolumn{1}{c}{FC5}&\multicolumn{1}{c}{FC6}&\multicolumn{1}{c}{FC7}&\multicolumn{1}{c}{FC8}&\multicolumn{1}{c}{FC9}&\multicolumn{1}{c}{d}\tabularnewline
\hline
1&&&&&&&&&0.78&$ 8.08$\tabularnewline
2&&&&&&&&&0.42&$ 0.90$\tabularnewline
3&&-0.54&&&&&&&&$ 3.36$\tabularnewline
4&&&0.88&&&&&&&$ 1.38$\tabularnewline
\multicolumn{11}{c}{Block 5-6}\tabularnewline\hline
5&&&&&&-0.71&&&&$ 0.63$\tabularnewline
6&&&&&0.78&&&&&$ 4.81$\tabularnewline\hline
7&&&&&0.64&&&&&$ 2.81$\tabularnewline
\multicolumn{11}{c}{Block 8-11}\tabularnewline\hline
8&&-0.56&&&0.35&&&&&$ 3.38$\tabularnewline
9&&-0.63&&&&&&&&$ 2.18$\tabularnewline
10&&-0.53&&-0.32&&&&&&$ 4.14$\tabularnewline
11&&-0.58&&&&&&&&$ 1.81$\tabularnewline\hline
12&&&&&0.33&&&-0.44&&$ 3.40$\tabularnewline
13&&-0.63&&&&-0.41&&&&$ 3.40$\tabularnewline
14&-0.35&&&&&&&-0.47&&$ 1.01$\tabularnewline
\multicolumn{11}{c}{Block 15-16}\tabularnewline\hline
15&&-0.52&0.64&&&&&&&$ 0.91$\tabularnewline
16&&-0.39&0.35&-0.43&&&&&&$ 3.89$\tabularnewline\hline
17&&&&-0.74&&&&&&$ 0.37$\tabularnewline
18&&&&&&-0.81&&&&$ 0.70$\tabularnewline
19&&-0.58&&&&&&&&$ 2.73$\tabularnewline
20&&&&&&&&&&$ 0.79$\tabularnewline
\multicolumn{11}{c}{Block 21-24}\tabularnewline\hline
21&-0.74&&&&&&&&&$-0.18$\tabularnewline
22&-0.84&&&&&&&&&$ 0.83$\tabularnewline
23&-0.48&&&&0.37&&-0.4&&&$ 2.10$\tabularnewline
24&-0.39&&&&&&-0.5&&&$ 3.96$\tabularnewline
\multicolumn{11}{c}{Block 25-27}\tabularnewline\hline
25&&&&-0.86&&&&&&$ 0.57$\tabularnewline
26&-0.39&&&-0.61&&&&&&$-1.34$\tabularnewline
27&-0.36&&&&&&&&&$ 1.52$\tabularnewline\hline
28&&&0.74&&&&&&&$ 1.91$\tabularnewline
29&&&&&&&&&&$ 1.63$\tabularnewline
30&&&&&&&&&&$ 0.67$\tabularnewline
\hline
\end{tabular}\end{center}
\end{table}

While ANOVA demonstrated that the 9-factor model was statistically
superior, the model fit statistics shown in Table \ref{tab:fit}
did not provide a clear identification of the number of factors.
While the 9-factor model is statistically significantly better
than all other models, there was not a significant improvement
from 6-factor model to the 7-factor model [$\chi^2(24)=32.79$,
$p=0.109$]. The 9-factor model was a significant improvement over
the 6-factor model [$\chi^2(69)=196$, $p<0.001$]. The 5-factor
model had superior RMSEA, CFI, and TLI statistics. While the
9-factor model minimized AIC, the 6-factor model minimized BIC. The ``knee'' in the Scree plot calculated using traditional EFA,
presented in the Supplemental Materials \cite{supp}, suggests 3 to 4 factors. A such,
after 3 factors are extracted, it is difficult to make a
definitive case for the number of factors. We will examine the 9-factor model
because it was the model identified as optimal using the chi-squared test, minimized AIC, and it
provides the greatest resolution of the structure of the instrument. Three- and five-factor MIRT
models are presented in the Supplemental Materials.

\begin{table}[!tbp]
\caption{\label{tab:fit} MIRT fit statistics}
\begin{center}
\begin{tabular}{cccccc}
Factors&AIC&BIC&RMSEA&TLI&CFI\\\hline
1&132,042&132,430&0.071&0.83&0.84\\
2&128,805&129,379&0.047&0.92&0.94\\
3&127,863&128,619&0.042&0.94&0.95\\
4&127,223&128,153&0.038&0.95&0.96\\
5&126,553&127,651&0.032&0.97&0.98\\
6&126,239&127,498&0.066&0.85&0.91\\
7&126,254&127,668&0.071&0.83&0.91\\
8&126,192&127,755&0.067&0.85&0.92\\
9&126,180&127,885&0.060&0.88&0.94\\
10&126,214&128,055&0.066&0.86&0.94\\\hline\hline
\end{tabular}\end{center}
\end{table}

Traditional EFA was also performed. For this analysis, the
criteria that the eigenvalue be greater than the mean eigenvalue
suggested a 7-factor solution, parallel analysis suggested a
6-factor solution, while an examination of the ``knee'' in the
Scree plot suggested 3-4 factors. Like other published Scree
plots, there was a rapid decline from 1-3 factors followed by a
long tail where additional factors each explained 2-4\% additional
variance. If Huffman and Heller's criteria for the retained
factors, which were required to explain 5-10\% of the variance,
was used \cite{huffman_what_1995}, only two factors would have
been retained. The 5-factor EFA solution is presented in the Supplemental Materials.
The 5-factor solution was very similar
to other published solutions with many items loading on the first
two factors as was also observed by Scott, Schumayer, and Gray
\cite{scott2012exploratory}. Exploratory methods, such as factor
analysis or cluster analysis, can identify structures correlated
by unexpected features. The items in the first two factors in
either the 5-factor EFA model in the Supplemental Material or in
Scott, Schumayer, and Gray do not seem strongly related by the
physical principles they test, which opens the possibility that
some other feature is causing the correlations which cause groups
of items to be identified as factors.

\subsection{Correlation Analysis}

\begin{figure}[htb]
\includegraphics[width=3in]{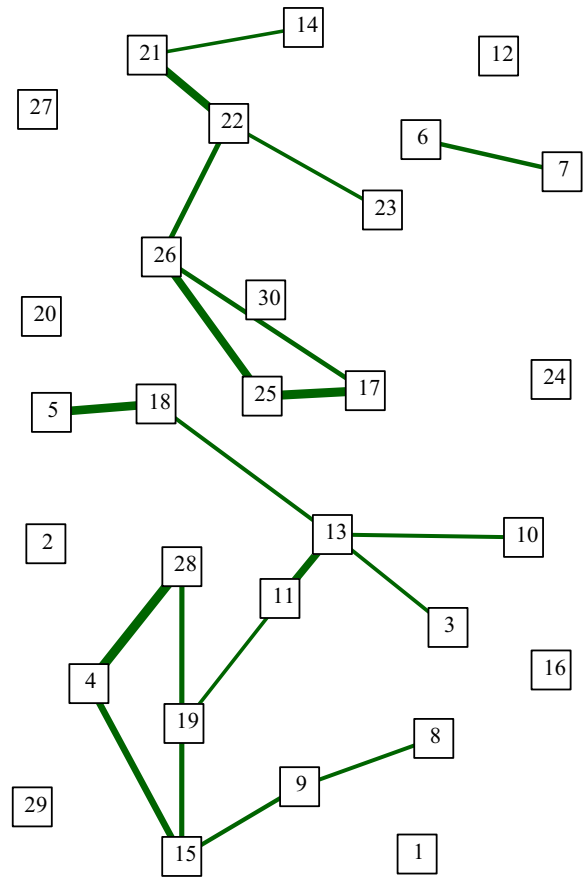}
\caption{\label{fig:corrfull} Correlation matrix for all FCI
items. Lines represent correlations with $|r|>0.3$. Line thickness
represents the size of the correlation. Solid (green) lines
represent positive correlations; dashed (red) lines negative
correlations. No negative correlations were present. }
\end{figure}

Factor analysis accomplished either traditionally or through MIRT
identifies combinations of items which vary together. Co-variation
of individual items can also be examined through correlation
analysis. The full FCI correlation matrix contains 900 entries
making it difficult to interpret; however, numerous visualizations
of the correlation matrix have been created. Figure
\ref{fig:corrfull} presents one such visualization of the FCI
correlation matrix. Solid lines (green) represent positive
correlations and dashed lines (red) represent negative
correlations. Only correlations greater than $0.3$ (Cohen's
criteria for medium effect size) are shown. No pair of questions
was negatively correlated with $|r|>0.3$ where $r$ is the
correlation coefficient and, therefore, there are no dashed lines
in the figure.

There are many potential sources of the correlations shown in Fig.
\ref{fig:corrfull}. Groups of highly correlated items often form
the elements of a factor with the highest loading; in some sense
they ``nucleate'' the factor. Some correlations may arise because
two items require similar physical principles for their solution
or that they elicit the same misconception. In previous factor
analysis, only these explanations of the factor structure have
been considered.

\begin{figure*}[htb]
\includegraphics[width=7in]{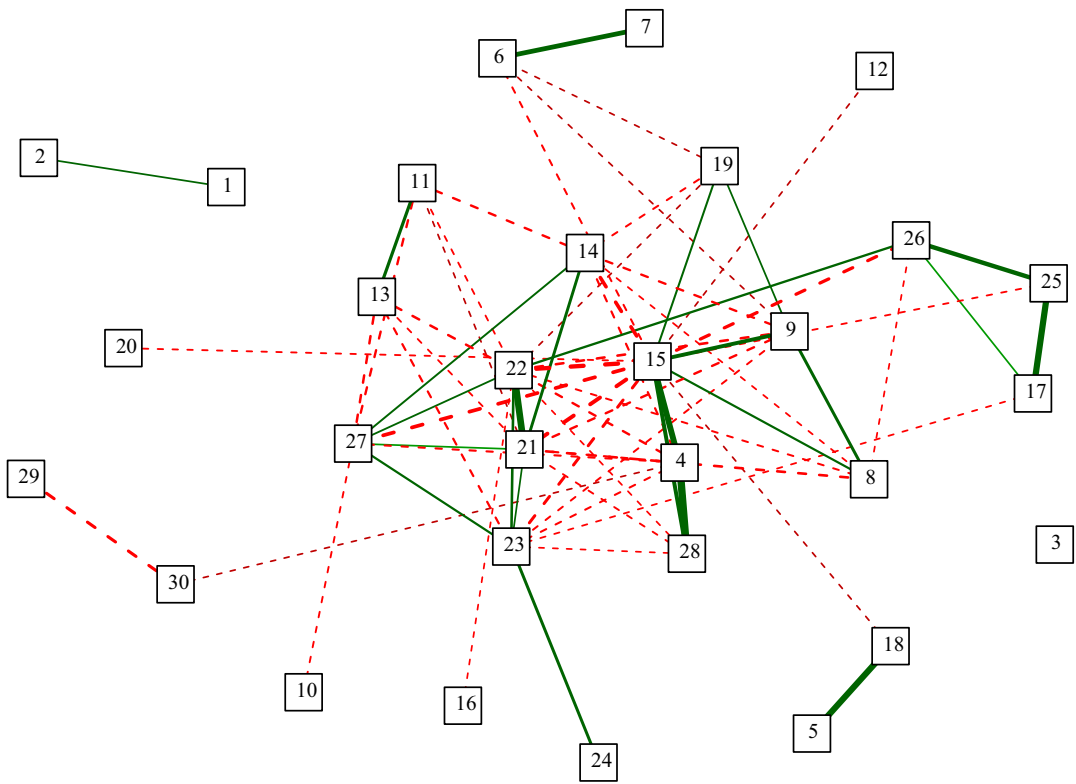}
\caption{\label{fig:corrpart} Partial correlation matrix for all
FCI problems correcting for total FCI score (only $|r|>0.1$
shown). Line thickness represents the size of the correlation.
Solid (green) lines represent positive correlations; dashed (red)
lines negative
correlations. } 
\end{figure*}

The FCI contains 4 groups of problems where each item in the group
share a common stem; we will call these groups ``problem blocks.''
The problem blocks have been identified in  Table \ref{tab:mirt-full}.
One additional group of items 25-27 does not share the same stem,
but items 26 and 27 explicitly refer to item 25. While blocking
the problems may shorten the reading time for the student, it can
also generate correlations between items that are not the result
of the physical properties required for their solution. If a
student misinterprets the stem, then this error will affect the
solution of each problem in the block. An error in an earlier item
in a block can cause errors in later items. Examination of Table
\ref{tab:mirt-full}
 shows that many of the largest factor loadings in individual factors
occur for problems in the same block; likewise, in Fig.
\ref{fig:corrfull} many of the most strongly correlated item pairs
are part of problem blocks. An examination of the physical
principles required to solve the strongly correlated blocked
problems does not suggest the level of commonality demonstrated by
the factor or correlation structure. As such, it seems likely that
at least some of the factor and correlation structure results for
the decision to use groups of problems with a common stem.

A second possible source of correlations not related to underlying
physical principles is correlation through total test score. The
FCI has repeatedly been shown to be an instrument with high
internal consistency as measured by Cronbach's alpha
\cite{lasry2011puzzling,traxler2017}. All correlations with
$|r|>0.3$ are positive in Fig. \ref{fig:corrfull}. Two problems
could be correlated because either only the strongest students
answer them correctly or only the weakest students answer them
incorrectly; they are correlated through the total test score. To
remove this effect, a partial correlation matrix controlling for
total test score was calculated as shown in Fig. \ref{fig:corrpart}.
Examination of Fig. \ref{fig:corrpart} shows that the problem
blocks \{8,9\}, \{21,22,23,24\}, and \{25,26\} still stand out as
highly correlated. Four other groups of questions emerge as
correlated \{5,18\}, \{6,7\}, \{17,25\}, and \{4,15,28\}. To
understand these groups, we will construct a model of the solution
to the FCI in the next section.

\subsection{A Theoretical Framework}
\label{sec:theory}

For over 50 years, social scientists have argued that research
shouldn't rely purely on exploratory techniques but rather that having
a robust theoretical framework is paramount to the determination
of model validity \cite{cronbach1955construct}. According to
Cronbach and Meele, there is no validity without an articulated
theory and it is, therefore, inappropriate to use only exploratory
techniques, such as EFA, on an instrument. Furthermore, EFA
results provide only information about the data itself and should
not be construed as providing genuine answers or solutions without
a theoretical core \cite{clark1995constructing}. Exploratory
methods generally identify some structure, and without a
framework, that structure may simply be the result of random
fluctuations in the data.

Hundreds of physicists have offered models of the structure of
introductory mechanics either through the production of textbooks,
scientific papers, or in their solution of introductory mechanics
problems. We sought to produce one such model that synthesized the
structure of introductory mechanics commonly presented in
textbooks with the statements found in expert solutions of FCI
problems. This resulted in a set of statements about introductory
mechanics shown in Table \ref{tab:theory}; the statements will be
called ``principles'' following Larkin {\it et al.}
\cite{larkin1980models}.  The principles were classified as
Definitions (DF), Laws (L), Corollaries (C), Results (R), Facts
(F), Lemmas (LM), and Reasoning (RS). Corollaries could be derived
from laws, results, and definitions but required some non-trivial
reasoning. A result, such as the constant acceleration kinematic
equations, was derived as a special case of the laws and
definitions. Knowledge of how the universe worked that did not
raise to the level of a law were called facts. Expert solutions
often contained specializations of the physical laws and
definitions to the individual problem; these special cases were
called lemmas. The FCI contains one item (item 19) which required
a unique piece of reasoning (RS1) in multiple expert solutions. To
solve the problem, one must argue if one quantity is constant and
another begins smaller than the first quantity and ends larger
than that quantity, then the quantities must be equal at some
point. Many of the principles in Table \ref{tab:theory} are
consistent with principles used in models of physics problem
solving proposed by Larkin {\it et al.} \cite{larkin1980models}
and Reif and Heller \cite{reif1982knowledge}.

To map out the subset of Newtonian mechanics tested by the FCI, a
careful solution of the FCI was collected from the lead instructor
who oversaw the course studied. Solutions were also collected from
faculty and graduate students in the research team. These
solutions were decomposed to the sentence level and each sentence
classified. These statements did not fully map out the higher
level structure of mechanics. Many lemmas provided qualitative
descriptions or specialization of more general principles not
found in the expert solutions. The general principles were
introduced based on the project team's understanding of Newtonian
physics. With only a small sample, it became obvious that a
complete set of secondary principles (lemmas) would be very long
and not particularly useful, but that the existing lemmas fit well
into a well-established general structure of mechanics. The
definitions, laws, and results, which form Newtonian mechanics as measured by the FCI,
were then completed producing the model shown in Table
\ref{tab:theory}. The table also shows the higher level principle
from which a subsidiary principle can be derived and the FCI items
whose solution requires the principle.

\begin{table*}
\caption{\label{tab:theory} Theoretical model of Newtonian
mechanics as tested by the FCI. Principles in bold were included
in the optimal model 3 fitting the reduced FCI.}
\begin{center}
\begin{tabular}{llll}
Label&Derived&FCI\#&\multicolumn{1}{c}{Principle}\\
&From&&\\\hline
\multicolumn{4}{c}{{\bf Kinematics}}\\
{\bf DF1}&&19,20&Definition of velocity ($\vec v = d\vec r/dt$)\\
{\bf DF2}&&&Definition of acceleration ($\vec a = d\vec v/dt$)\\
{\bf R1}&&&Trajectory $\vec a=$constant ($\vec r(t)=\vec r_0+\vec v_0 t+\frac12\vec at^2$)\\
{\bf R2}&&&Velocity $\vec a=$constant ($\vec v(t)=\vec v_0 t+\vec at$)\\
{\bf C1}&DF1&6,7&Instantaneous velocity is tangent to the trajectory.\\
{\bf C2}&DF2&5,18&Objects moving in a curved trajectory will experience centripetal acceleration.\\
{\bf C3}&R1&&1D trajectory $a=$constant, ($x(t)=x_0+v_0t+\frac12 at^2$)\\
{\bf C4}&R2&&1D velocity $a=$constant, ($v(t)=v_0+at$)\\
LM1&DF1&14&If two objects move together, they have the same initial velocity when separated.\\
LM2&R1&2&Motion may be separated along orthogonal axes.\\
LM3&C3&2&If motion is one-dimensional and $a=0$, then $d=vt$.\\
LM4&R2&3,22,26&Objects under constant acceleration with $\vec a$ parallel to $\vec v$ speed up.\\
LM5&R2&27&Objects under constant acceleration with $\vec a$ opposite to $\vec v$ slow down.\\
LM6&R1&12,14,21&Objects under constant acceleration with some initial velocity perpendicular\\
&&&\hspace{0.2in}to the acceleration travel in a parabolic arc.\\
LM7&DF2&20&If velocity is constant, then acceleration is zero.\\
LM11&C3&1,2&If the accelerations and initial velocities are equal, objects travel \\
&&&\hspace{0.2in}the same distance in the same time.\\
\multicolumn{4}{c}{{\bf Dynamics}}\\
{\bf DF3}&&26&The net force is the vector sum of the forces (forces add as vectors).\\
{\bf L1}&&6,7,8,10,17,23,24,25&Newton's 1st law\\
{\bf L2}&&5,18,26,27&Newton's 2nd law\\
{\bf L3}&&4,15,16,28&Newton's 3rd law\\
LM8&L2&21&Constant force produces constant acceleration.\\
LM9&L2&8,21&If the force only has one component, an object accelerates in that direction.\\
LM10&DF3&17,25&If the net force is zero and only two forces are exerted on the object,\\
&&&\hspace{0.2in}they must be equal but opposite.\\
\multicolumn{4}{c}{{\bf Properties of Forces}}\\
{\bf L4}&&1,2,3,5,11,12,13,14&Objects near the earth's surface experience a constant downward\\
&&17,18,29,30&force/acceleration of gravity.\\
{\bf F1}&&11,29&An object in contact with a surface experiences a normal force.\\
{\bf F2}&&11,13,18,30&An object does not necessarily experience a force in the direction of motion.\\
{\bf F3}&&3,29&Air pressure does not exert a net downward force.\\
{\bf F4}&&30&The wind can exert a force on an object.\\
{\bf F5}&&1,3&Air resistance is negligible for a compact object moving a short distance.\\
{\bf F6}&&3&The force of gravity is approximately constant near the earth's surface.\\
F7&&27&Objects that slide across a surface experience a force of friction opposite motion.\\
\multicolumn{4}{c}{{\bf Other}}\\
DF4&&&Magnitude of vector ($|\vec A|=\sqrt{A_x^2+A_y^2+A_z^2}$)\\
C5&DF4&9&Triangle inequality\\
RS1&&19&If one quantity is constant and another quantity is smaller at one time and \\
&&&\hspace{0.2in} larger at another time, then the two quantities must be equal at some time.\\
\end{tabular}
\end{center}
\end{table*}

The model in Table \ref{tab:theory} represents a preliminary model
for understanding solutions of the FCI. It does not contain any
representation of student misconceptions. The set of lemmas would
almost certainly change somewhat if a different set of expert
solutions were used. Some parts of this model would be agreed upon
by most experts, DF1, DF2, L2, and L3 for example. However, it is
doubtful that a group of experts would agree on all elements. For
example, it might be argued that Newton's 1st law is unnecessary
because it can be derived from Newton's 2nd law and kinematics.
Also, it might be argued that separate principles for
one-dimensional kinematics (C3 and C4) and three-dimensional kinematics (R1 and
R2) are unnecessary.

We note that the fact F2 might be considered specifically addressing
the motion implies force misconception. It was present in most
expert solutions to eliminate specific distractors. We will find
that its inclusion improves the model and future work may identify
other facts that allow common misconceptions to be added to FCI
models.

There were some additional minor decisions made to produce the
model in
 Table \ref{tab:theory}. Item 17 has a distractor that requires the application
 of F3 (net downward force of the air); no expert solution included this principle and
 it was not included in the model of item 18. LM4 and LM5 were written for
 general three-dimensional motion and are marked as derived from R2. Items 26 and 27, which
 use these lemmas, are one-dimensional problems. As the lemmas are folded into the principles
 they are derived from to produce model 3 below, the items using these lemmas will be appropriately
 distributed to one- or three-dimensional kinematic principles. Item 18 was coded with a centripetal
 acceleration implying a force in the direction of the tension force; this item could have also
 been coded by introducing the tension force as an additional fact. The correlation with item 5 and
 the lack of any additional items using a tension force caused the selection of this coding. Law L4 and fact F6
 both involve a constant force of gravity near the earth's surface. Fact F6 was introduced because FCI item 3 seems
 to require the student to explicitly reason that the force of gravity does not change much over the
 height of a single story building.

It may be impossible to produce a definitive expert model of the
FCI; however, the model in Table \ref{tab:theory} can be used to
understand some of the correlations in Fig. \ref{fig:corrpart}.
MIRT will allow the construction of the subset of the model that best
represents the divisions in student thinking for the population of
students studied.

\subsection{Reduced Exploratory Factor Analysis}

The theoretical framework in Table \ref{tab:theory} provides an
explanation for some of the remaining strong correlations in Fig.
\ref{fig:corrpart} which were not explained by the block structure
of the FCI. Items 4, 15, and 28 all require only L3 (Newton's 3rd
law) for their solution. Items 17 and 25 share both L1 and LM10,
items 5 and 18 share L2, L4, and C2, and items 6 and 7 share C1 and L1.
While Newton's 3rd law plays a central role in Newtonian mechanics
and, therefore, one would expect it to be repeated multiple times
in the FCI, the repetition of the other combinations of principles
is difficult to support theoretically as combinations somehow
central to mechanics and thus deserving special focus. The FCI
authors did not discuss the choice to include the problem pairs
\{5,18\}, \{6,7\} and \{17,25\} and therefore it seems likely the
inclusion of these pairs of very similar problems was accidental.
The inclusion of these problems does not affect the ability of the
instrument to measure an overall force concept beyond the
reduction of the breadth of the instrument; however, the
repetition of these problems does impact the correlation and
exploratory factor structure. Figure \ref{fig:corrpart} shows the
scores on these pairs of problems are highly correlated and these
pairs make up the strongest loading in factors FC4, FC5, and FC6. It
seems likely that the strong correlations of the pairs was part of
the reason these factors were extracted and that the factor
structure could be significantly modified by removing one problem
of each pair and inserting problems that repeated a different set
of principles. As such, any general conclusion drawn from the
existence factors FC4, FC5, or FC6 about the structure of knowledge
of Newtonian mechanics is suspect.

These factors based largely on pairs of questions also serve to
explain the relatively universal structure of the Scree plots
reported in this and other works. The Scree plots reported all
decrease strongly from 1-3 factors and then the amount of variance
explained by additional factors decreases rapidly. If a factor is
mostly capturing the co-variance of two items, the amount of
variance it can explain will be small.

\begin{figure}[htb]
\includegraphics[width=3in]{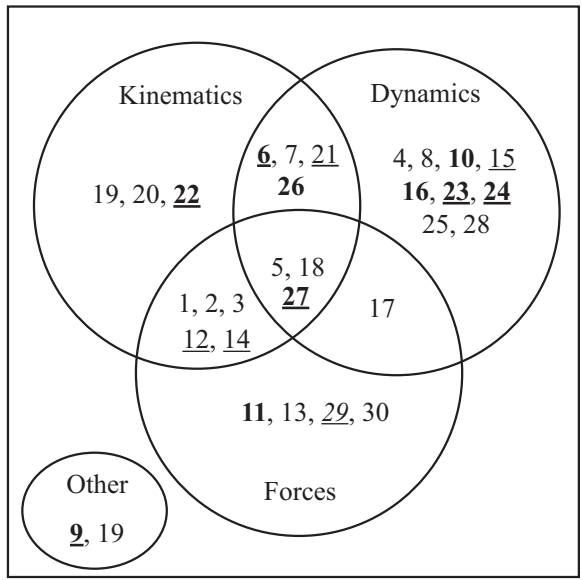}
\caption{\label{fig:venn} Venn Diagram of the distribution of
problems in the the reduced FCI. Items in bold are the blocked
items removed from the analysis. Underlined items are items
identified as unfair to men or women by Traxler {\it et
al.} \cite{traxler2017} (Item 29 was identified as fair but
unreliable). }
\end{figure}

With these observations, much of the original factor structure
identified by EFA appears to be either a result of the block
structure of the FCI or of repeated problems with very similar
solution structure. Removing all but the first problem in each
problem block and the second of the repeated problem pairs
produces a reduced 18-item instrument. Because item 6 was removed
due to blocking, item 7 was retained. The optimal MIRT model for
this set of problems is shown in Table \ref{tab:efa18}; 6 factors
were optimal.

\begin{table}[!tbp]
\caption{\label{tab:efa18} Exploratory factor analysis for the
reduced FCI (varimax rotation). Only loadings greater than $0.3$
are shown.}
\begin{center}
\begin{tabular}{lllllllr}
\hline\hline
\multicolumn{1}{l}{FCI \#}&\multicolumn{1}{c}{FC1}&\multicolumn{1}{c}{FC2}&\multicolumn{1}{c}{FC3}&\multicolumn{1}{c}{FC4}&\multicolumn{1}{c}{FC5}&\multicolumn{1}{c}{FC6}&\multicolumn{1}{c}{d}\tabularnewline
\hline
1&&&0.31&&&0.75&$ 6.25$\tabularnewline
2&&&&&&0.65&$ 0.93$\tabularnewline
3&&-0.60&&&&&$ 3.35$\tabularnewline
4&&&0.86&&&&$ 1.15$\tabularnewline
5&&-0.34&&-0.32&&&$ 0.32$\tabularnewline
7&&&&-0.33&-0.43&&$ 2.28$\tabularnewline
8&&-0.48&0.33&&-0.36&&$ 3.06$\tabularnewline
12&&&&-0.55&&&$ 2.82$\tabularnewline
13&&-0.70&&&&&$ 3.02$\tabularnewline
14&&&&-0.67&&&$ 0.65$\tabularnewline
15&&-0.47&0.64&&&&$ 1.04$\tabularnewline
17&-0.33&&&&&&$ 0.18$\tabularnewline
19&&-0.68&&&&&$ 2.90$\tabularnewline
20&-0.41&&&&&&$ 0.76$\tabularnewline
21&&&&-0.62&&&$-0.47$\tabularnewline
28&&&0.81&&&&$ 1.92$\tabularnewline
29&&&&&&&$ 1.67$\tabularnewline
30&&-0.35&&-0.33&&&$ 0.51$\tabularnewline
\hline
\end{tabular}\end{center}
\end{table}

Examination of Table \ref{tab:efa18} shows some factors that map
onto the theoretical model of mechanics. The problems have been
placed in a Venn Diagram based on the general classification in
Table \ref{tab:theory}. All FCI items have been included in the diagram.
Items removed to eliminate blocking are bolded. Unfair items identified by
Traxler {\it et al.} \cite{traxler2017} are underlined; these will be discussed later.
Few factors contain loadings that are
localized to individual regions of the Venn Diagram. There are
also loadings that cannot be supported theoretically. Factor FC3
contains the Newton's 3rd law items, but it also loads on items 1
and 8 which have nothing to do with Newton's 3rd law. Likewise,
item 15, which requires only Newton's 3rd law for its solution,
also loads strongly on FC2. It is also difficult to understand why
item 17 (force in elevator) and item 20 (blocks moving at
different speeds) form Factor FC1. It is unclear if correlations through the
overall difficulty of the item could explain some of the unexpected structure.

\subsection{Constrained MIRT}

\begin{table*}[!tbp]
\caption{\label{tab:mirt-decomp} Hierarchical MIRT modelling. The
$\chi^2$ test determines whether the models are statistically
different; if so, it measures the improvement of the superior
model over the inferior model.}
\begin{center}
\begin{tabular}{ccccccc}
\hline\hline
Original&Transformation&Transformed&AIC&BIC&Chi-Squared&Superior\\
Model&&Model&&&Test&Model\\\hline
0&&&91,067&91,668&\\
1&Remove all lemmas.&2&90,943&91,518&$\chi^2(4)=116, p<0.001$&2\\
2&Remove RS1.&3&90,920&91,488&$\chi^2(1)=21, p<0.001$&3\\
3&Combine DF3 with L2.&4&90,942&90,488&&3\\
3&Combine L1 with L2.&5&90,929&91,491&$\chi^2(1)=11, p=0.001$&3\\
3&Combine C3 with R1; C4 with R2.&6&90,991&91,553&$\chi^2(1)=73, p<0.001$&3\\
3&Remove F2.&7&90,941&91,490&$\chi^2(3)=27, p<0.001$&3\\
3&Remove F5.&8&90,944&91,499&$\chi^2(2)=28, p<0.001$&3\\
3&Remove F6.&9&90,929&91,491&$\chi^2(1)=11, p=0.001$&3\\
3&Replace L1 with L2 and DF2.&10&90,933&91,521&$\chi^2(3)=7.5,
p=0.058$&3\\\hline\hline
\end{tabular}\end{center}
\end{table*}

EFA failed to produce factors that could be reliably mapped onto
recognizable subdivisions of Newtonian mechanics. MIRT provides an
alternate avenue to explore how the students think about Newtonian
mechanics. The ${\bf a_j}$ parameter matrix can be constrained so
that parameters that should not theoretically affect a factor are
zero. For example, if the model of Newtonian mechanics in Table
\ref{tab:theory} was used as the basis for a constrained MIRT
model, then the factor representing DF1, $a_{DF1}$, could be
constrained to be zero except for items 19 and 20. A sequence of
models was constructed from the model in Table \ref{tab:theory}
which made small modifications to the original model; ANOVA was
then used to compare the models. For this analysis, only the first
problem in a problem block was retained as before; groups
of similar problems \{5,18\}, \{6,7\}, \{17,25\}, and \{4,15,28\} were
also retained. Because constrained MIRT is not exploratory, the
correlations of these items will not unduly influence the
analysis. The 20-item problem  set analyzed in this section was
then: 1, 2, 3, 4, 5, 7, 8, 12, 13, 14, 15, 17, 18, 19, 20, 21, 25,
28, 29, and 30.

The first model (model 1) included all the principles introduced
in Table \ref{tab:theory} which were not eliminated by removing
blocked items. F7 and C5 were eliminated when blocked items were
removed. The FCI has strong internal consistency and most items
are positively correlated. To separate a general facility with
Newtonian mechanics from a specific facility with one of the
principles, an additional factor was added that loaded on all
items. The fit statistics of model 1 are shown in Table
\ref{tab:mirt-decomp}. Because the parameter matrix was so sparse,
fit parameters such as CFI, RMSEA, and TLI could not be
calculated. Models can still be compared statistically with a
chi-squared test or by comparing AIC and BIC. Each transformation
in Table \ref{tab:mirt-decomp} modified the original model to the
transformed model. A chi-squared test determined whether the
models were statistically different. Model 4 did not change the
number of degrees of freedom from model 3; therefore, a
chi-squared test could not be performed; however, AIC and BIC
could be compared. Some models removed a principle from a previous
model. Other models combined two principles. For example, in model
5 all items that loaded on either L1 or L2 were set to load on
only L2. These models do not exhaust the set of available models,
but represented a set of models where a theoretical case could be
made for each change.

Each lemma represented a qualitative interpretation of some more
general principle or a special case of a general principle. To
determine if the lemmas were important to the understanding of the
pattern of answers, model 2 was constructed which removed all
lemmas and replaced them with the more general principle from
which they were derived. Model 2 was a significant improvement
over model 1 and, therefore, the answering pattern for this sample
could be understood without the lemmas. Model 3 removed the
crossing reasoning step, RS1; this also improved model fit. RS1
was used only in a subset of expert responses; other experts
simply observed that two of the interval lengths were comparable.
Model 4 explored whether the vector addition of forces could be
viewed as a part of Newton's 2nd law; model 3 was a significant
improvement over model 4. These students answer Newton's 2nd law
questions and addition of forces questions with different
facility. Combining Newton's 1st law and Newton's 2nd law to form
model 5 also did not improve model fit over model 3. A second
model that eliminated Newton's 1st law, model 10, replaced L1 with
L2 (Newton's 2nd law) and DF2 (the definition of acceleration).
This model was not statistically superior  to model 3 and the
model increased both AIC and BIC. As such, L1 was retained as a
separate entity. Combining C3 and C4 representing one-dimensional
kinematics into R1 and R2 representing three-dimensional
kinematics to form model 6 did not improve model fit over model 3.
Fact F2 addresses a common misconception; removing F2 from model 3
to form model 7 did not improve model fit. Finally, facts F5 (air
resistance is negligible) and F6 (gravity is approximately
constant) are additional pieces of information about mechanics;
however, their use was only required to eliminate distractors and
they were not used by some experts who solved the problem without
considering the distractors. Neither model 8 which eliminated F5
from model 3 nor model 9 which eliminated F6 from model 3 improved
model fit. As such, model 3 which contains all of Newton's 3 laws
with a separate definition of the addition of forces, leaves 1D
and 3D kinematics separate, and contains Facts 1-6 represented the
best model of students' responses to the FCI. Interestingly, model
3 is probably closest to the model presented in traditional
textbooks. Model 3 was also the model which minimized both AIC and
BIC.

\def\blk#1{{\bf #1}} 
\def\unrel#1{\underline{\it #1}} 
\def\biased#1{\underline{#1}}
\def\blkb#1{\underline{\bf #1}} 

\begin{table}
\caption{\label{tab:theoryred} Principles included in the optimal
model of the FCI, model 3. Items in bold are the blocked items
removed from the analysis. Underlined items are items identified
as unfair to men or women by Traxler {\it et al.}
\cite{traxler2017} (Item 29 was identified as fair but
unreliable).}
\begin{tabular}{llc}\hline
Label&Derived&FCI\#\\
&From&\\\hline \multicolumn{3}{c}{{\bf Kinematics}}\\\hline
DF1&&\biased{14},19,20\\
DF2&&20\\
R1&&2,\biased{12},\biased{14},\biased{21}\\
R2&&3,\blkb{22}\\
C1&DF1&\blkb{6},7\\
C2&DF2&5,18\\
C3&R1&1,2\\
C4&R2&\blk{26},\blkb{27}\\\hline
\multicolumn{3}{c}{{\bf Dynamics}}\\\hline
DF3&&17,25,\blk{26}\\
L1&&\blkb{6},7,8,\blk{10},17,\blkb{23},\blkb{24},25\\
L2&&5,8,18,\biased{21},\blk{26},\blkb{27}\\
L3&&4,\biased{15},\blk{16},28\\\hline
\multicolumn{3}{c}{{\bf Properties of Forces}}\\\hline
L4&&1,2,3,5,\blk{11},\biased{12},13,\biased{14},17,18,\unrel{29},30\\
F1&&\blk{11},\unrel{29}\\
F2&&\blk{11},13,18,30\\
F3&&3,\unrel{29}\\
F4&&30\\
F5&&1,3\\
F6&&3\\
F7&&\blkb{27}\\\hline
\multicolumn{3}{c}{{\bf Other}}\\\hline
C5&&\blkb{9}\\\hline
\end{tabular}
\end{table}

Model 3 with the transformations applied is presented in Table
\ref{tab:theoryred}. The factor loadings for model 3 are presented
in Table \ref{tab:mirt-opt}.  For this model, the $a_0$
coefficient represents the factor that was loaded on all items
representing the overall discrimination of the item. To allow
comparison with the more intuitive 2PL model, an effective
difficulty, $b_j$, is calculated $b_j=-d_j/a_{0j}$. The larger
$b_j$ the lower the probability the students will answer the item
correctly; the 2PL probability function is shown in Eqn.
\ref{eqn:2pl}.  The ``mirt'' package does not report normalized
latent variables; the standard deviation of the each latent
variable has been absorbed into the $a_j$ coefficient. Therefore,
the $a_j$ coefficient represents the change in log odds if the
latent trait increases by one standard deviation.

\begin{table*}[!tbp]
\caption{\label{tab:mirt-opt} Optimal MIRT model. The number in
parenthesis is the discrimination, $a_{jk}$, for the item. $a_0$
is the discrimination for a factor loaded on all items and $b$ is
the difficulty of the item.}
\begin{center}
\begin{tabular}{llll}
\hline\hline
$\#$&Model&$a_0$&$b$\\\hline
1&C3($1.04\pm0.30$)  L4($-0.30\pm0.19$)  F5($0.09\pm0.11$)&$2.25\pm0.44$&$-3.30\pm0.22$\\
2&R1($0.06\pm0.05$)  C3($0.48\pm0.12$)  L4($0.09\pm0.05$)&$1.05\pm0.07$&$-0.86\pm0.06$\\
3&R2($0.13\pm0.09$)  L4($0.02\pm0.12$)  F3($0.14\pm0.09$) F5($0.13\pm0.09$) F6($0.15\pm0.09$)&$1.65\pm0.19$&$-2.50\pm0.12$\\
4&L3($2.37\pm0.29$)&$1.88\pm0.19$&$-0.72\pm0.05$\\
5&C2($0.64\pm0.15$)  L2($0.51\pm0.10$)  L4($0.50\pm0.11$)&$1.49\pm0.13$&$-0.38\pm0.05$\\
7&C1($0.16\pm0.09$)  L1($0.01\pm0.05$)&$0.64\pm0.06$&$-3.42\pm0.26$\\
8&L1($-0.27\pm0.08$)  L2($-0.30\pm0.09$)&$1.41\pm0.12$&$-2.18\pm0.09$\\
12&R1($0.55\pm0.08$)  L4($0.18\pm0.07$)&$0.75\pm0.07$&$-3.73\pm0.31$\\
13&L4($0.29\pm0.10$)  F2($0.27\pm0.09$)&$2.36\pm0.17$&$-1.34\pm0.05$\\
14&DF1($0.22\pm0.08$)  R1($1.03\pm0.15$)  L4($0.31\pm0.08$)&$0.78\pm0.07$&$-0.99\pm0.09$\\
15&L3($0.79\pm0.05$)&$0.87\pm0.06$&$-0.78\pm0.07$\\
17&DF3($0.70\pm0.13$)  L1($0.60\pm0.12$) L4($0.14\pm0.06$)&$1.64\pm0.13$&$-0.18\pm0.04$\\
18&C2($0.65\pm0.14$)  L2($0.58\pm0.11$) L4($0.50\pm0.11$) F2($0.27\pm0.09$)&$1.71\pm0.13$&$-0.31\pm0.04$\\
19&DF1($0.16\pm0.08$)&$1.28\pm0.08$&$-2.04\pm0.09$\\
20&DF1($0.44\pm0.12$)  DF2($0.23\pm0.10$)&$1.12\pm0.09$&$-0.83\pm0.05$\\
21&R1($0.82\pm0.10$)  L2($0.28\pm0.07$)&$0.80\pm0.07$&$0.62\pm0.07$\\
25&DF3($0.70\pm0.13$)  L1($0.60\pm0.13$)&$1.91\pm0.16$&$-0.13\pm0.03$\\
28&L3($1.28\pm0.09$)&$1.70\pm0.09$&$-0.98\pm0.05$\\
29&L4($-0.10\pm0.09$)  F1($0.09\pm0.06$) F3($0.09\pm0.06$)&$0.17\pm0.06$&$-12.12\pm6.28$\\
30&L4($0.29\pm0.09$) F2($0.19\pm0.07$) F4($0.24\pm0.10$)&$1.11\pm0.08$&$-0.57\pm0.05$\\
\hline
\end{tabular}\end{center}
\end{table*}

\begin{figure*}[htb]
\includegraphics[width=6in]{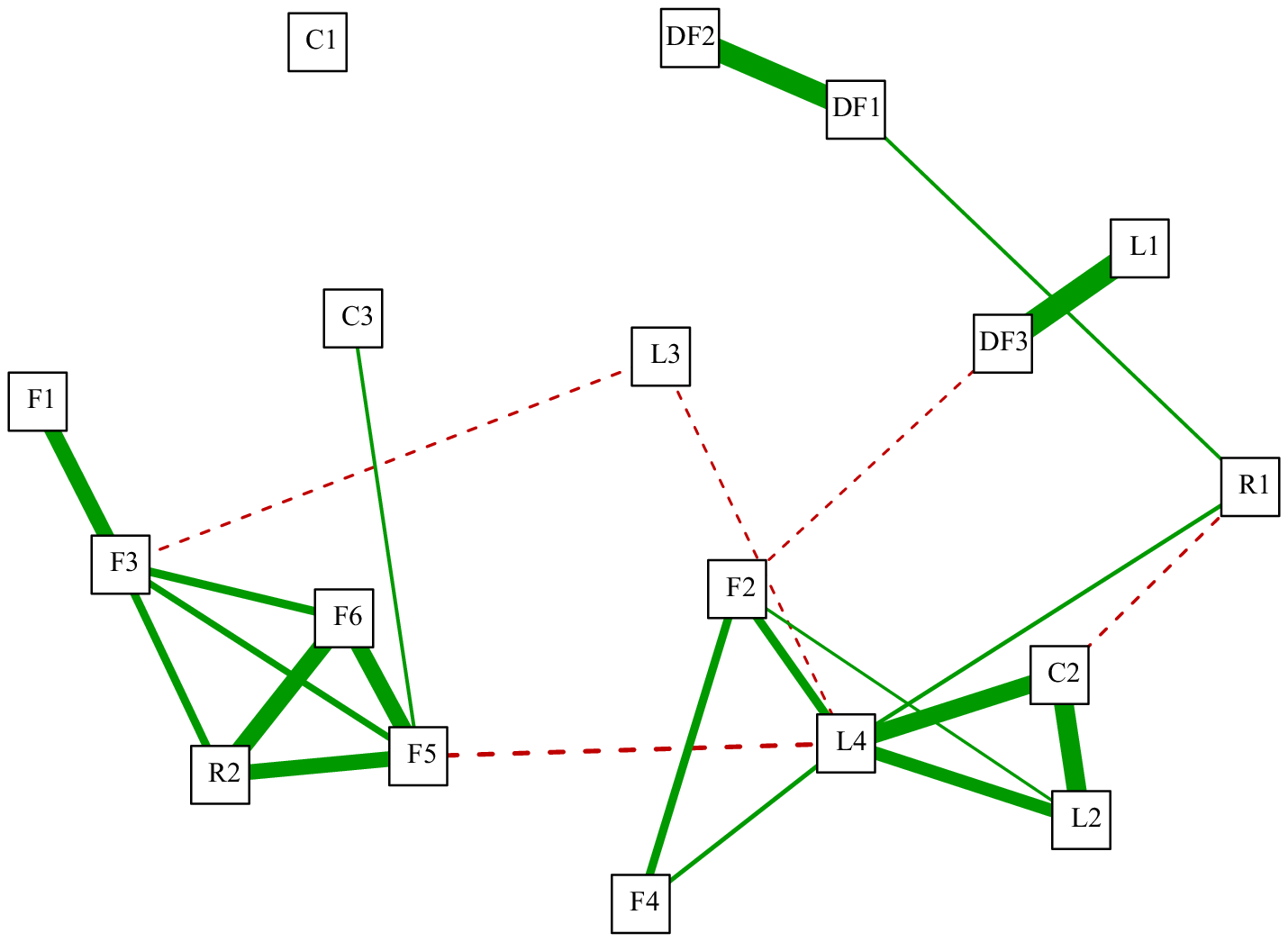}
\caption{\label{fig:corr-ability} Correlation matrix of student
ability using model 3. Lines represent correlations with
$|r|>0.15$. Line thickness represents the size of the correlation.
Solid (green) lines represent positive correlations; dashed (red)
lines negative correlations.  } 
\end{figure*}

Some alternate forms of the constrained analysis were also
performed. The optimal model in Table \ref{tab:mirt-opt} included
one factor that loaded on all problems; a factor capturing a
students overall facility with conceptual Newtonian mechanics. The
model with this factor (AIC=90,920, BIC=91,488) was a significant
improvement over the model without this overall factor
(AIC=94,442, BIC=94,881)[$\chi^2(20)=3562$, p$<$0.001]. The model
with this factor also had superior behavior in tests that compared
the optimal model to models where additional factors that damaged
the model had been introduced. For example, the addition of L3
(Newton's 3rd law) to item 1 produced a significantly less well
fitting model with the overall factor, but not without it. The
model without this overall factor is presented in the Supplemental
Materials \cite{supp}.

MIRT can estimate each student's ability on each trait. The
correlation matrix for the student ability by trait is shown in
Fig. \ref{fig:corr-ability}. For expert-like understanding, we
would expect a single overall ability to solve mechanics problems
and, therefore, that all correlations be equally strong.

\subsection{Comparison with Original FCI Model}

The FCI authors suggested a detailed structure for the FCI
dividing the test into 6 general categories and 23 fine-grained
principles (see Table I in \cite{fci}). The fine-grained
principles play the same role as the principles in the theoretical
model in Table \ref{tab:theory}. The FCI was revised in 1995; the
revised test included 3 new problems which were not categorized.
These items, revised FCI items 5, 18, and 30, will not be included
in this analysis.

Fitting a model implementing the structure suggested in the
original FCI paper on the set of items
1, 2, 3, 4, 7, 8, 12, 13, 14, 15, 17, 19, 20, 21, 25, 28, and 29 from the revised FCI produced a model
with AIC$=75,260$ and BIC=$75,453$. Using the constrained MIRT
model of the previous section on the more restricted problem set
produced a model with substantially better model fit
[AIC=$74,812$; BIC=$75,277$]. ANOVA showed that the constrained
MIRT model had significantly better model fit [$\chi^2(13)=474$,
$p<0.001$]. As such, while the model proposed by the FCI authors
captured their motivation as the creators of the instrument, model
3 produced a better fit for this student population.

\section{Discussion}

\label{sec:discussion} This study investigated five research
questions; they will be discussed in the order proposed.

{\it RQ1: What factor structure is extracted for the FCI by MIRT?
Is this structure consistent with the results of other factor
analysis?} MIRT identified a 9-factor solutions as optimal for the
full 30-item FCI. Other studies have identified 5-factor
\cite{scott2012exploratory} and 6-factor \cite{semak2017}
post-test models as optimal. It is possible that the larger sample
used in the present study combined with the strong incentives
given for correctly answering the items allowed this study to
resolve more detailed structure in the FCI. The 9-factor model,
while the best statistically based on nested $\chi^2$ tests, was
not the best model on all fit statistics (Table \ref{tab:fit}).
The fit statistics could also support the identification of either
a 5-factor or 6-factor model. All three of these studies
identified more factors than Huffman and Heller
\cite{huffman_what_1995}; however, this may have resulted from the
differing size and quality of the samples as well as the different
criteria used to select the optimal number of factors.

The factors extracted can also be compared. If the 9-factor
solution found in this study resulted because of superior
resolution of the factors, we would expect some of the factors in
the previously reported models to split to form the additional
factors in this study. Some commonality can be found between the
5-factor \cite{scott2012exploratory}, 6-factor \cite{semak2017},
and our 9-factor model. The groups of physically similar items
\{5,18\}, \{6,7\}, \{17,25\}, and \{4, 15, 16, 28\} do factor
together in all models, except that item 16 often does not factor
with the Newton's 3rd law group. The 5-factor model shows the same
tendency of blocked items to factor together that we saw in the
9-factor model; this effect was less pronounced in the 6-factor
model. All the factor models are difficult to support in terms of
the actual structure of the physical principles needed to solve
the problems shown in Table \ref{tab:theory}. As such, it is
difficult to support the proposition that EFA is providing
fundamental insights into the knowledge structure of physics
students as measured by the FCI.

{\it RQ2: Can parts of this factor structure be explained by
factors other than the structure of student knowledge of Newtonian
mechanics?}

Correlation analysis identified two non-physical sources of
relations between FCI items which could affect the factor
structures: correlations through the blocking of items into groups
and correlations through total score. The effect of blocking was
clear in Table \ref{tab:mirt-full} with most blocked questions
sharing the same factor with the exception of items 5-6. The
strong correlation of many blocked items can also be seen in the
overall correlation matrix (Fig. \ref{fig:corrfull}). Only the
first item in each group was retained; the non-physical
correlations created by blocking could not be corrected
statistically.  While the possible correlation of blocked items
seems relatively uncontroversial, we know of no previous research
that identifies it as a possible source of a non-physical
perturbation on the factor structure or other analysis. The
possible correlation between total score could be deduced through
the studies showing the FCI as a very internally consistent
instrument \cite{lasry2011puzzling, traxler2017}  as well as
Huffman and Heller's identification of the FCI as a single-factor
instrument \cite{huffman_what_1995}. This internal consistency is
clearly demonstrated in Fig. \ref{fig:corrfull} showing all
correlations are positive. The possibility of the difficulty of an
item impacting the factor structure was discussed briefly by
Scott, Schumayer, and Gray \cite{scott2012exploratory}.

The correlations through overall test score were removed by
calculating a partial correlation matrix (Fig. \ref{fig:corrpart})
which continued to show the effect of problem blocking and
revealed a third source of correlation. There were four groups of
items in the FCI which are answerable using very similar physical
principles. One group, items requiring Newton's 3rd law for their
solution, was expected. This group forms one of the factors in
each published analysis
\cite{scott2012exploratory,scott2015,semak2017} except Huffman and
Heller \cite{huffman_what_1995}. The other three groups do not
seem to represent special combinations of reasoning particularly
important to understanding mechanics and the repetition of these
principles seems likely to be accidental. These groups \{5,18\},
\{6,7\}, and \{17,25\} had large factor loadings in the same
factor in all published models. It seems likely that the
repetition of these blocks artificially influenced the factor
structure; many other equally important combinations of physical
reasoning could have been repeated.

{\it RQ3: If blocked items and repeated reasoning groups are
removed, is the resulting factor structure consistent with
Newtonian mechanics?} An EFA was also presented for a reduced set
of FCI items which removed all but the first item in each problem
block and removed the second item of the \{5,18\} and \{17,25\}
groups and the first item of the \{6,7\} group. This EFA found a
6-factor solution (Table \ref{tab:efa18}); however, the factors
make little physical sense. Factor 1 mixed a Newton's 1st law
problem involving an elevator with the analysis of two plots with
zero acceleration. Factor 2 contains a mixture of items including
Newton's 3rd law, one-dimensional constant acceleration, and a
position vs. time plot involving objects of constant velocity and
acceleration. Factor 3 includes three Newton's 3rd law items but
also two-dimensional zero acceleration motion and one-dimensional
motion under gravity. As such, factor analysis, once non-physical
correlations are removed, does not extract a factor structure
consistent with Newtonian mechanics. As the designers intended,
the FCI is a single-factor instrument
\cite{hestenes1995interpreting}. The reason for the coherence can
be seen in Fig. \ref{fig:venn} where many items test multiple
general domains.

{\it RQ4: Can theoretically constrained MIRT produce a model of
the physical constructs measured by the FCI? If so, what is the
optimal model of the FCI for this student population?}

Constrained MIRT allowed the exploration of a set of related
models grounded in the traditional theoretical framework of
Newtonian mechanics. This exploration showed, while expert
solutions to the FCI were cast in a number of lemmas which
converted the mathematical framework of mechanics to
language-based principles, that these were not needed to
understand the structure of student understanding. The optimal
model supported the differentiation in student thought between
Newton's 1st law and Newton's 2nd law as well as the difference
between one-dimensional and three-dimensional constant
acceleration kinematics. Facility with the vector addition of
forces was also shown to be distinct from facility with Newton's
2nd law.

Table \ref{tab:mirt-opt} shows the optimal MIRT model 3. The number in
parenthesis next to the principle label is the discrimination for
the principle. Because an overall factor loading on all items was
included, $a_0$, the discrimination, $a_{j>0}$, of the individual
principles represents the additional effect of the specific ability
over the student's general ability with Newtonian mechanics. Some
of the discrimination parameters are very small indicating that
the item does not require additional facility with the principle
over the student's general ability to answer FCI questions
correctly. Some discriminations are negative which may be a sign
of a problematic item. Items with only one strongly discriminating
principle might be claimed to be good marker items for the skill
represented by the principle. Items 1, 2, 12, 14, and 21 require
multiple principles but discriminate on one principle more
strongly than the others. These questions might be used to
characterize the students knowledge on the high discrimination
principle. Items 4, 15, 19, and 28 require only one principle, and
therefore could be used as a measure ability to perform this
principle; however, three of the four represent Newton's 3rd law.
Items 5, 17, 18, and 25 require multiple principles with
commensurate and large discriminations. These items measure
multiple abilities at the same time, but do not differentiate
between the abilities. Finally, a number of items have small
discrimination values for all principles: items 3, 7, 13, 29, and
30. These items do not contribute additional information about
specific abilities. Item 8 had negative discrimination; this may
indicate the item is not functioning correctly.

MIRT provides a new lens for examining physics evaluations. If
this lens proves valuable, it will suggest certain desirable
properties in future evaluations. First, the structure and number
of items should allow non-compensatory MIRT models to be fit to
extract item-level difficulty parameters. Second, each item should
provide additional information about some ability. Third, the
instrument should be invertible so that a linear combination of the scores on a
subset of items provides an estimate of the ability for a each
principle; thus giving practitioners a detailed characterization
of their learning outcomes.

MIRT can also be used to estimate the ability of each student to
answer each item. The correlations of these abilities were
presented in Fig. \ref{fig:corr-ability}. Because one factor was
loaded onto all items, these abilities represent that difference
between the students' general ability to solve a conceptual
mechanics question and his or her ability to apply a specific
principle. For a student with a fully developed expert
understanding of mechanics, we would expect their ability to apply
each principle to be equal, and therefore their difference in
ability to be zero. Fig. \ref{fig:corr-ability} shows multiple
principles with large correlations and large differences in the
strength of the correlation between different items. From this
diagram, we can infer that the students studied have differing but
correlated abilities with concepts of velocity and acceleration
(DF1,DF2), with Newton's 1st law (L1) and the addition of forces
(DF3), and with Newton's 2nd law (L2) and the law of gravitation
(L4). Additional instruction may be required to allow students to
fully integrate these concepts. MIRT, then, may also represent a
tool which can be used to probe the structure of knowledge and to
quantitatively characterize expert/novice differences and to
localize where additional integration of knowledge is needed.

{\it RQ5: Does the structure proposed by the FCI's authors provide
a superior description of the instrument to the optimal model
identified by MIRT?} The structure suggested by the authors of the
FCI \cite{fci} was also fit to the dataset and the result compared
to the optimal model 3 identified by MIRT. Model 3 outperformed
the model suggested with the publication of the FCI. As such, part
of the reason the published structure has not been recovered may
be that other models fit the FCI better. This seems unlikely to be
the primary reason for the mismatch between the proposed model and
model 3. Table \ref{tab:mirt-opt} and Fig. \ref{fig:venn} as well
as Hestenes and Halloun's insistence that the FCI measures a
single Newtonian force concept \cite{hestenes1995interpreting}
show that the instrument simply was not constructed to factor
well. There are very few items that use a single principle and
only Newton's 3rd law, not Newton's 1st or 2nd law, is used
independently and is repeated multiple times in the unblocked
model (Table \ref{tab:mirt-opt}). Most FCI items measure multiple
physical principles at once.

This work identified the blocking of items in the FCI as a source
of correlations not related to the student's ability to answer
conceptual physics questions. To eliminate these correlations, only the
first item in a problem block should be retained; as such,
items 6, 9, 10, 11, 16, 22, 23, 24, 26, and 27 were removed from the FCI
producing a 20-item version of the FCI. The model in Table \ref{tab:theoryred} can be used
to understand the effect of this reduction. The blocked items to be removed are shown in bold in both
Table \ref{tab:theoryred} and Fig. \ref{fig:venn}. Removing these items eliminated
principles F7 and C5 while reducing coverage of R2 and C4. In general, these
reductions still leave the coverage of the FCI intact although the elimination
of an explicit use of friction is a loss.

Traxler {\it et al.}
\cite{traxler2017} also suggested a reduced 19-item instrument
(including  FCI questions: 1, 2, 3, 4, 5, 7, 8, 10, 11, 13, 16,
17, 18, 19, 20, 25, 26, 28, and 30) to removed items with
reliability problems and to remove items unfair to either men or
women. The items removed to produce the 19-item instrument are underlined
in Table \ref{tab:theoryred} and Fig. \ref{fig:venn}. While this reduction removes
seven items from both kinematics and dynamics in Table \ref{tab:theoryred}, the coverage
of kinematics required more principles than dynamics. The removal of unfair items from
R1, R2, and C4 may substantially change the coverage of the instrument. Removing both
blocked and unfair items further reduces the coverage.

To produce a fair instrument while maintaining coverage, it may be
necessary to retain some blocked and unfair items but to balance
the degree and number of unfair items for both men and women.
Traxler {\it et al.} \cite{traxler2017} reported that two of the
removed items were unfair to men, items 9 and 15. If these two
items are retained as well as items 14 and 27, which were unfair
to women with similar Differential Item Functioning statistics,
the overall score should be gender fair. Blocked items 11 and 26
could also be retained to maintain coverage. Retaining these items
would increase coverage of some kinematic principles while
providing coverage of F7 and C5. This would leave a reduced
21-item FCI instrument containing: items 1, 2, 3, 4, 5, 7, 8 , 9,
11, 13, 14, 15, 17, 18, 19, 20, 25, 26, 27, 28, and 30. Blocked
items 10 and 16 were removed because there was sufficient coverage
of the principles required for their solution.

\section{Limitations}
This work was performed with a single sample drawn from a single
institution. Additional studies are necessary to determine if the
conclusions are general. The sample was analyzed in aggregate;
additional analysis is needed to determine if the results apply to
all student sub-groups. The analysis did not consider the role of
misconceptions; an extended theoretical model including
misconceptions should also be tested.

\section{Implications}
This worked showed a theoretical model of introductory mechanics
could be useful in understanding the results of conceptual
inventories. Such models can be constructed for other conceptual
areas of physics and could form a basic tool for understanding the
detailed results of PER instructional innovations. The constrained
MIRT analysis technique allowed the fine-grained exploration of
the constructs measured by the FCI and may be a powerful tool for
improving our understanding of student knowledge.  EFA did not
produce a factor structure that was useful in understanding the
FCI and it is likely that purely exploratory tools may not yield
generalizable results. Part of the reason for the failure of EFA
was correlations produced by the blocks of questions in the FCI.
The practice of using blocks of questions with the same stem may
make PER instruments difficult to interpret statistically and
should be discontinued.

This work showed that if all blocked items identified as problematic because of
correlations produced by blocking and all items identified as unfair or unreliable
by Traxler {\it et al.} \cite{traxler2017} are removed that the coverage of kinematics of
the modified FCI is reduced. This work proposed a 21-item reduced FCI to maintain
coverage while balancing unfair items; to have to decide between coverage and fairness is
unacceptable. While this 21-item instrument could be used for the near future, the identification
of unfair items and blocked items as problematic in addition to the lack of coherent sub-scales suggest that
it is time to revisit the construction of the FCI and to modernize it to remove some of the
difficulties identified in recent research.

\section{Future Work}
This work will be extended to analyze other conceptual instruments
popular in PER including the FMCE \cite{thornton1998} and the CSEM
\cite{maloney2001}. The work will also be extended to determine if
the results are consistent between men and women and to determine
if this method can help in understanding the differences observed
in male and female performance on conceptual evaluations.

\section{Conclusions}
\label{sec:conclusions}

The work examined the structure of the FCI with Multi-Dimensional
Item Response Theory (MIRT); first as an exploratory method and
then using constrained MIRT through the use of a theoretical model
of Newtonian mechanics. The exploratory analysis identified a
9-factor solution which showed some similarities to previously
published solutions. Further analysis showed many of the factors
in the 9-factor solution, and the previously published solutions,
could have resulted from the use of multiple problem blocks and
the repetition of physically similar items. Exploratory factor
analysis was repeated removing these correlated items; the
resulting 6-factor solution could not be reconciled with the
theoretical structure of Newtonian mechanics. Constrained MIRT was
then employed to determine the optimal model of the FCI for the
student population studied. The optimal model differentiated
between Newton's 1st and 2nd law; between Newton's 2nd law and the
principle of vector addition of forces; and between
one-dimensional and three-dimensional kinematics. The optimal
model identified by MIRT was substantially statistically superior
to the original model proposed by the authors of the FCI.

\begin{acknowledgments}
This work was supported in part by the National Science Foundation
as part of the evaluation of improved learning for the Physics
Teacher Education Coalition, PHY-0108787.
\end{acknowledgments}


\begin{thebibliography}{71}%
\makeatletter
\providecommand \@ifxundefined [1]{%
 \@ifx{#1\undefined}
}%
\providecommand \@ifnum [1]{%
 \ifnum #1\expandafter \@firstoftwo
 \else \expandafter \@secondoftwo
 \fi
}%
\providecommand \@ifx [1]{%
 \ifx #1\expandafter \@firstoftwo
 \else \expandafter \@secondoftwo
 \fi
}%
\providecommand \natexlab [1]{#1}%
\providecommand \enquote  [1]{``#1''}%
\providecommand \bibnamefont  [1]{#1}%
\providecommand \bibfnamefont [1]{#1}%
\providecommand \citenamefont [1]{#1}%
\providecommand \href@noop [0]{\@secondoftwo}%
\providecommand \href [0]{\begingroup \@sanitize@url \@href}%
\providecommand \@href[1]{\@@startlink{#1}\@@href}%
\providecommand \@@href[1]{\endgroup#1\@@endlink}%
\providecommand \@sanitize@url [0]{\catcode `\\12\catcode
`\$12\catcode
  `\&12\catcode `\#12\catcode `\^12\catcode `\_12\catcode `\%12\relax}%
\providecommand \@@startlink[1]{}%
\providecommand \@@endlink[0]{}%
\providecommand \url  [0]{\begingroup\@sanitize@url \@url }%
\providecommand \@url [1]{\endgroup\@href {#1}{\urlprefix }}%
\providecommand \urlprefix  [0]{URL }%
\providecommand \Eprint [0]{\href }%
\providecommand \doibase [0]{http://dx.doi.org/}%
\providecommand \selectlanguage [0]{\@gobble}%
\providecommand \bibinfo  [0]{\@secondoftwo}%
\providecommand \bibfield  [0]{\@secondoftwo}%
\providecommand \translation [1]{[#1]}%
\providecommand \BibitemOpen [0]{}%
\providecommand \bibitemStop [0]{}%
\providecommand \bibitemNoStop [0]{.\EOS\space}%
\providecommand \EOS [0]{\spacefactor3000\relax}%
\providecommand \BibitemShut  [1]{\csname bibitem#1\endcsname}%
\let\auto@bib@innerbib\@empty
\bibitem [{\citenamefont {Hestenes}\ \emph {et~al.}(1992)\citenamefont
  {Hestenes}, \citenamefont {Wells},\ and\ \citenamefont {Swackhamer}}]{fci}%
  \BibitemOpen
  \bibfield  {author} {\bibinfo {author} {\bibfnamefont {D.}~\bibnamefont
  {Hestenes}}, \bibinfo {author} {\bibfnamefont {M.}~\bibnamefont {Wells}}, \
  and\ \bibinfo {author} {\bibfnamefont {G.}~\bibnamefont {Swackhamer}},\
  }\bibfield  {title} {\enquote {\bibinfo {title} {{F}orce {C}oncept
  {I}nventory},}\ }\href@noop {} {\bibfield  {journal} {\bibinfo  {journal}
  {Phys. Teach.}\ }\textbf {\bibinfo {volume} {30}},\ \bibinfo {pages}
  {141--158} (\bibinfo {year} {1992})}\BibitemShut {NoStop}%
\bibitem [{\citenamefont {Hake}(1998)}]{hake1998}%
  \BibitemOpen
  \bibfield  {author} {\bibinfo {author} {\bibfnamefont {R.R.}\ \bibnamefont
  {Hake}},\ }\bibfield  {title} {\enquote {\bibinfo {title}
  {Interactive-engagement versus traditional methods: {A} six-thousand-student
  survey of mechanics test data for introductory physics courses},}\
  }\href@noop {} {\bibfield  {journal} {\bibinfo  {journal} {Am. J. Phys.}\
  }\textbf {\bibinfo {volume} {66}},\ \bibinfo {pages} {64--74} (\bibinfo
  {year} {1998})}\BibitemShut {NoStop}%
\bibitem [{\citenamefont {Thornton}\ and\ \citenamefont
  {Sokoloff}(1998)}]{thornton1998}%
  \BibitemOpen
  \bibfield  {author} {\bibinfo {author} {\bibfnamefont {R.K.}\ \bibnamefont
  {Thornton}}\ and\ \bibinfo {author} {\bibfnamefont {D.R.}\ \bibnamefont
  {Sokoloff}},\ }\bibfield  {title} {\enquote {\bibinfo {title} {Assessing
  student learning of {N}ewton's laws: The {F}orce and {M}otion {C}onceptual
  {E}valuation and the evaluation of active learning laboratory and lecture
  curricula},}\ }\href {\doibase 10.1119/1.18863} {\bibfield  {journal}
  {\bibinfo  {journal} {Am. J. Phys.}\ }\textbf {\bibinfo {volume} {66}},\
  \bibinfo {pages} {338--352} (\bibinfo {year} {1998})}\BibitemShut {NoStop}%
\bibitem [{\citenamefont {Maloney}\ \emph {et~al.}(2001)\citenamefont
  {Maloney}, \citenamefont {O'Kuma}, \citenamefont {Hieggelke},\ and\
  \citenamefont {Van~Huevelen}}]{maloney2001}%
  \BibitemOpen
  \bibfield  {author} {\bibinfo {author} {\bibfnamefont {D.P.}\ \bibnamefont
  {Maloney}}, \bibinfo {author} {\bibfnamefont {T.L.}\ \bibnamefont {O'Kuma}},
  \bibinfo {author} {\bibfnamefont {C.}~\bibnamefont {Hieggelke}}, \ and\
  \bibinfo {author} {\bibfnamefont {A.}~\bibnamefont {Van~Huevelen}},\
  }\bibfield  {title} {\enquote {\bibinfo {title} {Surveying students'
  conceptual knowledge of electricity and magnetism},}\ }\href {\doibase
  10.1119/1.1371296} {\bibfield  {journal} {\bibinfo  {journal} {Phys. Ed.
  Res., Am. J. Phys.}\ }\textbf {\bibinfo {volume} {69}},\ \bibinfo {pages}
  {S12--S23} (\bibinfo {year} {2001})}\BibitemShut {NoStop}%
\bibitem [{\citenamefont {Ding}\ \emph {et~al.}(2006)\citenamefont {Ding},
  \citenamefont {Chabay}, \citenamefont {Sherwood},\ and\ \citenamefont
  {Beichner}}]{ding2006}%
  \BibitemOpen
  \bibfield  {author} {\bibinfo {author} {\bibfnamefont {L.}~\bibnamefont
  {Ding}}, \bibinfo {author} {\bibfnamefont {R.}~\bibnamefont {Chabay}},
  \bibinfo {author} {\bibfnamefont {B.}~\bibnamefont {Sherwood}}, \ and\
  \bibinfo {author} {\bibfnamefont {R.}~\bibnamefont {Beichner}},\ }\bibfield
  {title} {\enquote {\bibinfo {title} {Evaluating an electricity and magnetism
  assessment tool: {B}rief {E}lectricity and {M}agnetism {A}ssessment},}\
  }\href@noop {} {\bibfield  {journal} {\bibinfo  {journal} {Phys. Rev. Phys.
  Educ. Res.}\ }\textbf {\bibinfo {volume} {2}},\ \bibinfo {pages} {010105}
  (\bibinfo {year} {2006})}\BibitemShut {NoStop}%
\bibitem [{\citenamefont {Docktor}\ and\ \citenamefont
  {Mestre}(2014)}]{docktor2014synthesis}%
  \BibitemOpen
  \bibfield  {author} {\bibinfo {author} {\bibfnamefont {J.L.}\ \bibnamefont
  {Docktor}}\ and\ \bibinfo {author} {\bibfnamefont {J.P.}\ \bibnamefont
  {Mestre}},\ }\bibfield  {title} {\enquote {\bibinfo {title} {Synthesis of
  discipline-based education research in physics},}\ }\href@noop {} {\bibfield
  {journal} {\bibinfo  {journal} {Phys. Rev. Phys. Educ. Res.}\ }\textbf
  {\bibinfo {volume} {10}},\ \bibinfo {pages} {020119} (\bibinfo {year}
  {2014})}\BibitemShut {NoStop}%
\bibitem [{\citenamefont {Huffman}\ and\ \citenamefont
  {Heller}(1995)}]{huffman_what_1995}%
  \BibitemOpen
  \bibfield  {author} {\bibinfo {author} {\bibfnamefont {D.}~\bibnamefont
  {Huffman}}\ and\ \bibinfo {author} {\bibfnamefont {P.}~\bibnamefont
  {Heller}},\ }\bibfield  {title} {\enquote {\bibinfo {title} {What does the
  {F}orce {C}oncept {I}nventory actually measure?}}\ }\href {\doibase
  10.1119/1.2344171} {\bibfield  {journal} {\bibinfo  {journal} {Phys. Teach.}\
  }\textbf {\bibinfo {volume} {33}},\ \bibinfo {pages} {138} (\bibinfo {year}
  {1995})}\BibitemShut {NoStop}%
\bibitem [{\citenamefont {Scott}\ \emph {et~al.}(2012)\citenamefont {Scott},
  \citenamefont {Schumayer},\ and\ \citenamefont
  {Gray}}]{scott2012exploratory}%
  \BibitemOpen
  \bibfield  {author} {\bibinfo {author} {\bibfnamefont {T.F.}\ \bibnamefont
  {Scott}}, \bibinfo {author} {\bibfnamefont {D.}~\bibnamefont {Schumayer}}, \
  and\ \bibinfo {author} {\bibfnamefont {A.R.}\ \bibnamefont {Gray}},\
  }\bibfield  {title} {\enquote {\bibinfo {title} {Exploratory factor analysis
  of a {F}orce {C}oncept {I}nventory data set},}\ }\href@noop {} {\bibfield
  {journal} {\bibinfo  {journal} {Phys. Rev. Phys. Educ. Res.}\ }\textbf
  {\bibinfo {volume} {8}},\ \bibinfo {pages} {020105} (\bibinfo {year}
  {2012})}\BibitemShut {NoStop}%
\bibitem [{\citenamefont {Semak}\ \emph {et~al.}(2017)\citenamefont {Semak},
  \citenamefont {Dietz}, \citenamefont {Pearson},\ and\ \citenamefont
  {Willis}}]{semak2017}%
  \BibitemOpen
  \bibfield  {author} {\bibinfo {author} {\bibfnamefont {M.R.}\ \bibnamefont
  {Semak}}, \bibinfo {author} {\bibfnamefont {R.D.}\ \bibnamefont {Dietz}},
  \bibinfo {author} {\bibfnamefont {R.H.}\ \bibnamefont {Pearson}}, \ and\
  \bibinfo {author} {\bibfnamefont {C.W.}\ \bibnamefont {Willis}},\ }\bibfield
  {title} {\enquote {\bibinfo {title} {Examining evolving performance on the
  {F}orce {C}oncept {I}nventory using factor analysis},}\ }\href {\doibase
  10.1103/PhysRevPhysEducRes.13.010103} {\bibfield  {journal} {\bibinfo
  {journal} {Phys. Rev. Phys. Educ. Res.}\ }\textbf {\bibinfo {volume} {13}},\
  \bibinfo {pages} {010103} (\bibinfo {year} {2017})}\BibitemShut {NoStop}%
\bibitem [{\citenamefont {Brewe}\ \emph {et~al.}(2016)\citenamefont {Brewe},
  \citenamefont {Bruun},\ and\ \citenamefont {Bearden}}]{brewe_using_2016}%
  \BibitemOpen
  \bibfield  {author} {\bibinfo {author} {\bibfnamefont {E.}~\bibnamefont
  {Brewe}}, \bibinfo {author} {\bibfnamefont {J.}~\bibnamefont {Bruun}}, \ and\
  \bibinfo {author} {\bibfnamefont {I.G.}\ \bibnamefont {Bearden}},\ }\bibfield
   {title} {\enquote {\bibinfo {title} {Using module analysis for multiple
  choice responses: {A} new method applied to {Force} {Concept} {Inventory}
  data},}\ }\href {\doibase 10.1103/PhysRevPhysEducRes.12.020131} {\bibfield
  {journal} {\bibinfo  {journal} {Phys. Rev. Phys. Educ. Res.}\ }\textbf
  {\bibinfo {volume} {12}},\ \bibinfo {pages} {020131} (\bibinfo {year}
  {2016})}\BibitemShut {NoStop}%
\bibitem [{\citenamefont {Springuel}\ \emph {et~al.}(2007)\citenamefont
  {Springuel}, \citenamefont {Wittmann},\ and\ \citenamefont
  {Thompson}}]{springuel2007applying}%
  \BibitemOpen
  \bibfield  {author} {\bibinfo {author} {\bibfnamefont {R.P.}\ \bibnamefont
  {Springuel}}, \bibinfo {author} {\bibfnamefont {M.C.}\ \bibnamefont
  {Wittmann}}, \ and\ \bibinfo {author} {\bibfnamefont {J.R.}\ \bibnamefont
  {Thompson}},\ }\bibfield  {title} {\enquote {\bibinfo {title} {Applying
  clustering to statistical analysis of student reasoning about two-dimensional
  kinematics},}\ }\href@noop {} {\bibfield  {journal} {\bibinfo  {journal}
  {Phys. Ed. Res., Am. J. Phys.}\ }\textbf {\bibinfo {volume} {3}},\ \bibinfo
  {pages} {020107} (\bibinfo {year} {2007})}\BibitemShut {NoStop}%
\bibitem [{\citenamefont {Stewart}\ \emph {et~al.}(2012)\citenamefont
  {Stewart}, \citenamefont {Miller}, \citenamefont {Audo},\ and\ \citenamefont
  {Stewart}}]{stewart2012}%
  \BibitemOpen
  \bibfield  {author} {\bibinfo {author} {\bibfnamefont {J.}~\bibnamefont
  {Stewart}}, \bibinfo {author} {\bibfnamefont {M.}~\bibnamefont {Miller}},
  \bibinfo {author} {\bibfnamefont {C.}~\bibnamefont {Audo}}, \ and\ \bibinfo
  {author} {\bibfnamefont {G.}~\bibnamefont {Stewart}},\ }\bibfield  {title}
  {\enquote {\bibinfo {title} {Using cluster analysis to identify patterns in
  students' responses to contextually different conceptual problems},}\ }\href
  {\doibase 10.1103/PhysRevSTPER.8.020112} {\bibfield  {journal} {\bibinfo
  {journal} {Phys. Rev. Phys. Educ. Res.}\ }\textbf {\bibinfo {volume} {8}},\
  \bibinfo {pages} {020112} (\bibinfo {year} {2012})}\BibitemShut {NoStop}%
\bibitem [{\citenamefont {Wang}\ and\ \citenamefont
  {Bao}(2010)}]{wang_analyzing_2010}%
  \BibitemOpen
  \bibfield  {author} {\bibinfo {author} {\bibfnamefont {J.}~\bibnamefont
  {Wang}}\ and\ \bibinfo {author} {\bibfnamefont {L.}~\bibnamefont {Bao}},\
  }\bibfield  {title} {\enquote {\bibinfo {title} {Analyzing {F}orce {C}oncept
  {I}nventory with item response theory},}\ }\href {\doibase 10.1119/1.3443565}
  {\bibfield  {journal} {\bibinfo  {journal} {Am. J. Phys.}\ }\textbf {\bibinfo
  {volume} {78}},\ \bibinfo {pages} {1064--1070} (\bibinfo {year}
  {2010})}\BibitemShut {NoStop}%
\bibitem [{\citenamefont {Planinic}\ \emph {et~al.}(2010)\citenamefont
  {Planinic}, \citenamefont {Ivanjek},\ and\ \citenamefont
  {Susac}}]{planinic2010}%
  \BibitemOpen
  \bibfield  {author} {\bibinfo {author} {\bibfnamefont {M.}~\bibnamefont
  {Planinic}}, \bibinfo {author} {\bibfnamefont {L.}~\bibnamefont {Ivanjek}}, \
  and\ \bibinfo {author} {\bibfnamefont {A.}~\bibnamefont {Susac}},\ }\bibfield
   {title} {\enquote {\bibinfo {title} {Rasch model based analysis of the
  {F}orce {C}oncept {I}nventory},}\ }\href {\doibase
  10.1103/PhysRevSTPER.6.010103} {\bibfield  {journal} {\bibinfo  {journal}
  {Phys. Rev. Phys. Educ. Res.}\ }\textbf {\bibinfo {volume} {6}},\ \bibinfo
  {pages} {010103} (\bibinfo {year} {2010})}\BibitemShut {NoStop}%
\bibitem [{\citenamefont {Osborn~Popp}\ \emph {et~al.}(2011)\citenamefont
  {Osborn~Popp}, \citenamefont {Meltzer},\ and\ \citenamefont
  {Megowan-Romanowicz}}]{popp2011}%
  \BibitemOpen
  \bibfield  {author} {\bibinfo {author} {\bibfnamefont {S.}~\bibnamefont
  {Osborn~Popp}}, \bibinfo {author} {\bibfnamefont {D.}~\bibnamefont
  {Meltzer}}, \ and\ \bibinfo {author} {\bibfnamefont {M.C.}\ \bibnamefont
  {Megowan-Romanowicz}},\ }\bibfield  {title} {\enquote {\bibinfo {title} {Is
  the {Force} {Concept} {Inventory} biased? {I}nvestigating differential item
  functioning on a test of conceptual learning in physics},}\ }in\ \href
  {http://www.aera.org} {\emph {\bibinfo {booktitle} {2011 {American}
  {Educational} {Research} {Association} {Conference}}}}\ (\bibinfo
  {publisher} {American Education Research Association},\ \bibinfo {address}
  {Washington, DC},\ \bibinfo {year} {2011})\BibitemShut {NoStop}%
\bibitem [{\citenamefont {Traxler}\ \emph {et~al.}(2018)\citenamefont
  {Traxler}, \citenamefont {Henderson}, \citenamefont {Stewart}, \citenamefont
  {Stewart}, \citenamefont {Papak},\ and\ \citenamefont
  {Lindell}}]{traxler2017}%
  \BibitemOpen
  \bibfield  {author} {\bibinfo {author} {\bibfnamefont {A.}~\bibnamefont
  {Traxler}}, \bibinfo {author} {\bibfnamefont {R.}~\bibnamefont {Henderson}},
  \bibinfo {author} {\bibfnamefont {J.}~\bibnamefont {Stewart}}, \bibinfo
  {author} {\bibfnamefont {G.}~\bibnamefont {Stewart}}, \bibinfo {author}
  {\bibfnamefont {A.}~\bibnamefont {Papak}}, \ and\ \bibinfo {author}
  {\bibfnamefont {R.}~\bibnamefont {Lindell}},\ }\bibfield  {title} {\enquote
  {\bibinfo {title} {Gender fairness within the {F}orce {C}oncept
  {I}nventory},}\ }\href {\doibase 10.1103/PhysRevPhysEducRes.14.010103}
  {\bibfield  {journal} {\bibinfo  {journal} {Phys. Rev. Phys. Educ. Res.}\
  }\textbf {\bibinfo {volume} {14}},\ \bibinfo {pages} {010103} (\bibinfo
  {year} {2018})}\BibitemShut {NoStop}%
\bibitem [{\citenamefont {Morris}\ \emph {et~al.}(2006)\citenamefont {Morris},
  \citenamefont {Branum-Martin}, \citenamefont {Harshman}, \citenamefont
  {Baker}, \citenamefont {Mazur}, \citenamefont {Dutta}, \citenamefont
  {Mzoughi},\ and\ \citenamefont {McCauley}}]{morris2006testing}%
  \BibitemOpen
  \bibfield  {author} {\bibinfo {author} {\bibfnamefont {G.A.}\ \bibnamefont
  {Morris}}, \bibinfo {author} {\bibfnamefont {L.}~\bibnamefont
  {Branum-Martin}}, \bibinfo {author} {\bibfnamefont {N.}~\bibnamefont
  {Harshman}}, \bibinfo {author} {\bibfnamefont {S.D.}\ \bibnamefont {Baker}},
  \bibinfo {author} {\bibfnamefont {E.}~\bibnamefont {Mazur}}, \bibinfo
  {author} {\bibfnamefont {S.}~\bibnamefont {Dutta}}, \bibinfo {author}
  {\bibfnamefont {T.}~\bibnamefont {Mzoughi}}, \ and\ \bibinfo {author}
  {\bibfnamefont {V.}~\bibnamefont {McCauley}},\ }\bibfield  {title} {\enquote
  {\bibinfo {title} {Testing the test: {I}tem response curves and test
  quality},}\ }\href@noop {} {\bibfield  {journal} {\bibinfo  {journal} {Am. J.
  Phys.}\ }\textbf {\bibinfo {volume} {74}},\ \bibinfo {pages} {449--453}
  (\bibinfo {year} {2006})}\BibitemShut {NoStop}%
\bibitem [{\citenamefont {Morris}\ \emph {et~al.}(2012)\citenamefont {Morris},
  \citenamefont {Harshman}, \citenamefont {Branum-Martin}, \citenamefont
  {Mazur}, \citenamefont {Mzoughi},\ and\ \citenamefont
  {Baker}}]{morris2012item}%
  \BibitemOpen
  \bibfield  {author} {\bibinfo {author} {\bibfnamefont {G.A.}\ \bibnamefont
  {Morris}}, \bibinfo {author} {\bibfnamefont {N.}~\bibnamefont {Harshman}},
  \bibinfo {author} {\bibfnamefont {L.}~\bibnamefont {Branum-Martin}}, \bibinfo
  {author} {\bibfnamefont {E.}~\bibnamefont {Mazur}}, \bibinfo {author}
  {\bibfnamefont {T.}~\bibnamefont {Mzoughi}}, \ and\ \bibinfo {author}
  {\bibfnamefont {S.D.}\ \bibnamefont {Baker}},\ }\bibfield  {title} {\enquote
  {\bibinfo {title} {An item response curves analysis of the {F}orce {C}oncept
  {I}nventory},}\ }\href@noop {} {\bibfield  {journal} {\bibinfo  {journal}
  {Am. J. Phys.}\ }\textbf {\bibinfo {volume} {80}},\ \bibinfo {pages}
  {825--831} (\bibinfo {year} {2012})}\BibitemShut {NoStop}%
\bibitem [{\citenamefont {Bao}\ and\ \citenamefont {Redish}(2006)}]{bao2006}%
  \BibitemOpen
  \bibfield  {author} {\bibinfo {author} {\bibfnamefont {L.}~\bibnamefont
  {Bao}}\ and\ \bibinfo {author} {\bibfnamefont {E.F.}\ \bibnamefont
  {Redish}},\ }\bibfield  {title} {\enquote {\bibinfo {title} {Model analysis:
  {R}epresenting and assessing the dynamics of student learning},}\ }\href
  {\doibase 10.1103/PhysRevSTPER.2.010103} {\bibfield  {journal} {\bibinfo
  {journal} {Phys. Rev. Phys. Educ. Res.}\ }\textbf {\bibinfo {volume} {2}},\
  \bibinfo {pages} {010103} (\bibinfo {year} {2006})}\BibitemShut {NoStop}%
\bibitem [{\citenamefont {Ding}\ and\ \citenamefont
  {Beichner}(2009)}]{ding2009}%
  \BibitemOpen
  \bibfield  {author} {\bibinfo {author} {\bibfnamefont {L.}~\bibnamefont
  {Ding}}\ and\ \bibinfo {author} {\bibfnamefont {R.}~\bibnamefont
  {Beichner}},\ }\bibfield  {title} {\enquote {\bibinfo {title} {Approaches to
  data analysis of multiple-choice questions},}\ }\href {\doibase
  10.1103/PhysRevSTPER.5.020103} {\bibfield  {journal} {\bibinfo  {journal}
  {Phys. Rev. Phys. Educ. Res.}\ }\textbf {\bibinfo {volume} {5}},\ \bibinfo
  {pages} {020103} (\bibinfo {year} {2009})}\BibitemShut {NoStop}%
\bibitem [{\citenamefont {Lasry}\ \emph {et~al.}(2011)\citenamefont {Lasry},
  \citenamefont {Rosenfield}, \citenamefont {Dedic}, \citenamefont {Dahan},\
  and\ \citenamefont {Reshef}}]{lasry2011puzzling}%
  \BibitemOpen
  \bibfield  {author} {\bibinfo {author} {\bibfnamefont {N.}~\bibnamefont
  {Lasry}}, \bibinfo {author} {\bibfnamefont {S.}~\bibnamefont {Rosenfield}},
  \bibinfo {author} {\bibfnamefont {H.}~\bibnamefont {Dedic}}, \bibinfo
  {author} {\bibfnamefont {A.}~\bibnamefont {Dahan}}, \ and\ \bibinfo {author}
  {\bibfnamefont {O.}~\bibnamefont {Reshef}},\ }\bibfield  {title} {\enquote
  {\bibinfo {title} {The puzzling reliability of the {F}orce {C}oncept
  {I}nventory},}\ }\href@noop {} {\bibfield  {journal} {\bibinfo  {journal}
  {Am. J. Phys.}\ }\textbf {\bibinfo {volume} {79}},\ \bibinfo {pages}
  {909--912} (\bibinfo {year} {2011})}\BibitemShut {NoStop}%
\bibitem [{\citenamefont {Henderson}(2002)}]{henderson2002common}%
  \BibitemOpen
  \bibfield  {author} {\bibinfo {author} {\bibfnamefont {C.}~\bibnamefont
  {Henderson}},\ }\bibfield  {title} {\enquote {\bibinfo {title} {Common
  concerns about the {F}orce {C}oncept {I}nventory},}\ }\href@noop {}
  {\bibfield  {journal} {\bibinfo  {journal} {Phys. Teach.}\ }\textbf {\bibinfo
  {volume} {40}},\ \bibinfo {pages} {542--547} (\bibinfo {year}
  {2002})}\BibitemShut {NoStop}%
\bibitem [{\citenamefont {Jorion}\ \emph {et~al.}(2015)\citenamefont {Jorion},
  \citenamefont {Gane}, \citenamefont {James}, \citenamefont {Schroeder},
  \citenamefont {DiBello},\ and\ \citenamefont
  {Pellegrino}}]{jorion_analytic_2015}%
  \BibitemOpen
  \bibfield  {author} {\bibinfo {author} {\bibfnamefont {N.}~\bibnamefont
  {Jorion}}, \bibinfo {author} {\bibfnamefont {B.D.}\ \bibnamefont {Gane}},
  \bibinfo {author} {\bibfnamefont {K.}~\bibnamefont {James}}, \bibinfo
  {author} {\bibfnamefont {L.}~\bibnamefont {Schroeder}}, \bibinfo {author}
  {\bibfnamefont {L.V.}\ \bibnamefont {DiBello}}, \ and\ \bibinfo {author}
  {\bibfnamefont {J.W.}\ \bibnamefont {Pellegrino}},\ }\bibfield  {title}
  {\enquote {\bibinfo {title} {An analytic framework for evaluating the
  validity of concept inventory claims},}\ }\href {\doibase 10.1002/jee.20104}
  {\bibfield  {journal} {\bibinfo  {journal} {J. Eng. Educ.}\ }\textbf
  {\bibinfo {volume} {104}},\ \bibinfo {pages} {454--496} (\bibinfo {year}
  {2015})}\BibitemShut {NoStop}%
\bibitem [{\citenamefont {Scott}\ and\ \citenamefont
  {Schumayer}(2015)}]{scott2015}%
  \BibitemOpen
  \bibfield  {author} {\bibinfo {author} {\bibfnamefont {T.F.}\ \bibnamefont
  {Scott}}\ and\ \bibinfo {author} {\bibfnamefont {D.}~\bibnamefont
  {Schumayer}},\ }\bibfield  {title} {\enquote {\bibinfo {title} {Students'
  proficiency scores within multitrait item response theory},}\ }\href
  {\doibase 10.1103/PhysRevSTPER.11.020134} {\bibfield  {journal} {\bibinfo
  {journal} {Phys. Rev. Phys. Educ. Res.}\ }\textbf {\bibinfo {volume} {11}},\
  \bibinfo {pages} {020134} (\bibinfo {year} {2015})}\BibitemShut {NoStop}%
\bibitem [{\citenamefont {Hestenes}\ and\ \citenamefont
  {Halloun}(1995)}]{hestenes1995interpreting}%
  \BibitemOpen
  \bibfield  {author} {\bibinfo {author} {\bibfnamefont {D.}~\bibnamefont
  {Hestenes}}\ and\ \bibinfo {author} {\bibfnamefont {I.}~\bibnamefont
  {Halloun}},\ }\bibfield  {title} {\enquote {\bibinfo {title} {Interpreting
  the {F}orce {C}oncept {I}nventory: {A} response to {M}arch 1995 critique by
  {H}uffman and {H}eller},}\ }\href@noop {} {\bibfield  {journal} {\bibinfo
  {journal} {Phys. Teach.}\ }\textbf {\bibinfo {volume} {33}},\ \bibinfo
  {pages} {502--502} (\bibinfo {year} {1995})}\BibitemShut {NoStop}%
\bibitem [{\citenamefont {Heller}\ and\ \citenamefont
  {Huffman}(1995)}]{heller1995interpreting}%
  \BibitemOpen
  \bibfield  {author} {\bibinfo {author} {\bibfnamefont {P.}~\bibnamefont
  {Heller}}\ and\ \bibinfo {author} {\bibfnamefont {D.}~\bibnamefont
  {Huffman}},\ }\bibfield  {title} {\enquote {\bibinfo {title} {Interpreting
  the {F}orce {C}oncept {I}nventory: {A} reply to {H}estenes and {H}alloun},}\
  }\href@noop {} {\bibfield  {journal} {\bibinfo  {journal} {Phys. Teach.}\
  }\textbf {\bibinfo {volume} {33}},\ \bibinfo {pages} {503--503} (\bibinfo
  {year} {1995})}\BibitemShut {NoStop}%
\bibitem [{\citenamefont {Ramlo}(2008)}]{ramlo2008validity}%
  \BibitemOpen
  \bibfield  {author} {\bibinfo {author} {\bibfnamefont {S.}~\bibnamefont
  {Ramlo}},\ }\bibfield  {title} {\enquote {\bibinfo {title} {Validity and
  reliability of the {F}orce and {M}otion {C}onceptual {E}valuation},}\ }\href@noop {}
  {\bibfield  {journal} {\bibinfo  {journal} {Am. J. Phys.}\ }\textbf {\bibinfo
  {volume} {76}},\ \bibinfo {pages} {882--886} (\bibinfo {year}
  {2008})}\BibitemShut {NoStop}%
\bibitem [{\citenamefont {Han}\ \emph {et~al.}(2015)\citenamefont {Han},
  \citenamefont {Bao}, \citenamefont {Chen}, \citenamefont {Cai}, \citenamefont
  {Pi}, \citenamefont {Zhou}, \citenamefont {Tu},\ and\ \citenamefont
  {Koenig}}]{han2015}%
  \BibitemOpen
  \bibfield  {author} {\bibinfo {author} {\bibfnamefont {J.}~\bibnamefont
  {Han}}, \bibinfo {author} {\bibfnamefont {L.}~\bibnamefont {Bao}}, \bibinfo
  {author} {\bibfnamefont {L.}~\bibnamefont {Chen}}, \bibinfo {author}
  {\bibfnamefont {T.}~\bibnamefont {Cai}}, \bibinfo {author} {\bibfnamefont
  {Y.}~\bibnamefont {Pi}}, \bibinfo {author} {\bibfnamefont {S.}~\bibnamefont
  {Zhou}}, \bibinfo {author} {\bibfnamefont {Y.}~\bibnamefont {Tu}}, \ and\
  \bibinfo {author} {\bibfnamefont {K.}~\bibnamefont {Koenig}},\ }\bibfield
  {title} {\enquote {\bibinfo {title} {Dividing the {F}orce {C}oncept
  {I}nventory into two equivalent half-length tests},}\ }\href {\doibase
  10.1103/PhysRevSTPER.11.010112} {\bibfield  {journal} {\bibinfo  {journal}
  {Phys. Rev. Phys. Educ. Res.}\ }\textbf {\bibinfo {volume} {11}},\ \bibinfo
  {pages} {010112} (\bibinfo {year} {2015})}\BibitemShut {NoStop}%
\bibitem [{\citenamefont {Lee}\ \emph {et~al.}(2008)\citenamefont {Lee},
  \citenamefont {Palazzo}, \citenamefont {Warnakulasooriya},\ and\
  \citenamefont {Pritchard}}]{lee2008measuring}%
  \BibitemOpen
  \bibfield  {author} {\bibinfo {author} {\bibfnamefont {Y.}~\bibnamefont
  {Lee}}, \bibinfo {author} {\bibfnamefont {D.J.}\ \bibnamefont {Palazzo}},
  \bibinfo {author} {\bibfnamefont {R.}~\bibnamefont {Warnakulasooriya}}, \
  and\ \bibinfo {author} {\bibfnamefont {D.E.}\ \bibnamefont {Pritchard}},\
  }\bibfield  {title} {\enquote {\bibinfo {title} {Measuring student learning
  with item response theory},}\ }\href@noop {} {\bibfield  {journal} {\bibinfo
  {journal} {Phys. Ed. Res., Am. J. Phys.}\ }\textbf {\bibinfo {volume} {4}},\
  \bibinfo {pages} {010102} (\bibinfo {year} {2008})}\BibitemShut {NoStop}%
\bibitem [{\citenamefont {Gliner}(1989)}]{gliner1}%
  \BibitemOpen
  \bibfield  {author} {\bibinfo {author} {\bibfnamefont {G.S.}\ \bibnamefont
  {Gliner}},\ }\bibfield  {title} {\enquote {\bibinfo {title} {College
  students' organization of mathematics word problems in relation to success in
  problem solving},}\ }\href@noop {} {\bibfield  {journal} {\bibinfo  {journal}
  {School Sci. Math.}\ }\textbf {\bibinfo {volume} {89}},\ \bibinfo {pages}
  {392--404} (\bibinfo {year} {1989})}\BibitemShut {NoStop}%
\bibitem [{\citenamefont {Gliner}(1991)}]{gliner2}%
  \BibitemOpen
  \bibfield  {author} {\bibinfo {author} {\bibfnamefont {G.S.}\ \bibnamefont
  {Gliner}},\ }\bibfield  {title} {\enquote {\bibinfo {title} {College
  students' organization of mathematics word problems in terms of mathematical
  structure vs. surface structure},}\ }\href@noop {} {\bibfield  {journal}
  {\bibinfo  {journal} {School Sci. Math.}\ }\textbf {\bibinfo {volume} {91}},\
  \bibinfo {pages} {105--110} (\bibinfo {year} {1991})}\BibitemShut {NoStop}%
\bibitem [{\citenamefont {Eylon}\ and\ \citenamefont
  {Reif}(1984)}]{eylon1984effects}%
  \BibitemOpen
  \bibfield  {author} {\bibinfo {author} {\bibfnamefont {B.S.}\ \bibnamefont
  {Eylon}}\ and\ \bibinfo {author} {\bibfnamefont {F.}~\bibnamefont {Reif}},\
  }\bibfield  {title} {\enquote {\bibinfo {title} {Effects of knowledge
  organization on task performance},}\ }\href@noop {} {\bibfield  {journal}
  {\bibinfo  {journal} {Cognition Instruct.}\ }\textbf {\bibinfo {volume}
  {1}},\ \bibinfo {pages} {5--44} (\bibinfo {year} {1984})}\BibitemShut
  {NoStop}%
\bibitem [{\citenamefont {Chi}\ \emph {et~al.}(1981)\citenamefont {Chi},
  \citenamefont {Feltovich},\ and\ \citenamefont
  {Glaser}}]{chi1981categorization}%
  \BibitemOpen
  \bibfield  {author} {\bibinfo {author} {\bibfnamefont {M.T.H.}\ \bibnamefont
  {Chi}}, \bibinfo {author} {\bibfnamefont {P.J.}\ \bibnamefont {Feltovich}}, \
  and\ \bibinfo {author} {\bibfnamefont {R.}~\bibnamefont {Glaser}},\
  }\bibfield  {title} {\enquote {\bibinfo {title} {Categorization and
  representation of physics problems by experts and novices},}\ }\href@noop {}
  {\bibfield  {journal} {\bibinfo  {journal} {Cognitive Sci.}\ }\textbf
  {\bibinfo {volume} {5}},\ \bibinfo {pages} {121--152} (\bibinfo {year}
  {1981})}\BibitemShut {NoStop}%
\bibitem [{\citenamefont {Schoenfeld}\ and\ \citenamefont
  {Herrmann}(1982)}]{schoenfeld1982problem}%
  \BibitemOpen
  \bibfield  {author} {\bibinfo {author} {\bibfnamefont {A.H.}\ \bibnamefont
  {Schoenfeld}}\ and\ \bibinfo {author} {\bibfnamefont {D.J.}\ \bibnamefont
  {Herrmann}},\ }\bibfield  {title} {\enquote {\bibinfo {title} {Problem
  perception and knowledge structure in expert and novice mathematical problem
  solvers.}}\ }\href@noop {} {\bibfield  {journal} {\bibinfo  {journal} {J.
  Exp. Psychol. Learn.}\ }\textbf {\bibinfo {volume} {8}},\ \bibinfo {pages}
  {484} (\bibinfo {year} {1982})}\BibitemShut {NoStop}%
\bibitem [{\citenamefont {Reif}\ and\ \citenamefont
  {Heller}(1982)}]{reif1982knowledge}%
  \BibitemOpen
  \bibfield  {author} {\bibinfo {author} {\bibfnamefont {F.}~\bibnamefont
  {Reif}}\ and\ \bibinfo {author} {\bibfnamefont {J.I.}\ \bibnamefont
  {Heller}},\ }\bibfield  {title} {\enquote {\bibinfo {title} {Knowledge
  structure and problem solving in physics},}\ }\href@noop {} {\bibfield
  {journal} {\bibinfo  {journal} {Educ. Psychol.}\ }\textbf {\bibinfo {volume}
  {17}},\ \bibinfo {pages} {102--127} (\bibinfo {year} {1982})}\BibitemShut
  {NoStop}%
\bibitem [{\citenamefont {Beatty}\ and\ \citenamefont
  {Gerace}(2002)}]{beatty2002probing}%
  \BibitemOpen
  \bibfield  {author} {\bibinfo {author} {\bibfnamefont {I.D.}\ \bibnamefont
  {Beatty}}\ and\ \bibinfo {author} {\bibfnamefont {W.J.}\ \bibnamefont
  {Gerace}},\ }\bibfield  {title} {\enquote {\bibinfo {title} {Probing physics
  students' conceptual knowledge structures through term association},}\
  }\href@noop {} {\bibfield  {journal} {\bibinfo  {journal} {Am. J. Phys.}\
  }\textbf {\bibinfo {volume} {70}},\ \bibinfo {pages} {750--758} (\bibinfo
  {year} {2002})}\BibitemShut {NoStop}%
\bibitem [{\citenamefont {Miller}(1956)}]{miller1956magical}%
  \BibitemOpen
  \bibfield  {author} {\bibinfo {author} {\bibfnamefont {George~A}\
  \bibnamefont {Miller}},\ }\bibfield  {title} {\enquote {\bibinfo {title} {The
  magical number seven, plus or minus two: {S}ome limits on our capacity for
  processing information.}}\ }\href@noop {} {\bibfield  {journal} {\bibinfo
  {journal} {Psychol. Rev.}\ }\textbf {\bibinfo {volume} {63}},\ \bibinfo
  {pages} {81} (\bibinfo {year} {1956})}\BibitemShut {NoStop}%
\bibitem [{\citenamefont {Clement}(1982)}]{clement1982students}%
  \BibitemOpen
  \bibfield  {author} {\bibinfo {author} {\bibfnamefont {J.}~\bibnamefont
  {Clement}},\ }\bibfield  {title} {\enquote {\bibinfo {title} {Students'
  preconceptions in introductory mechanics},}\ }\href@noop {} {\bibfield
  {journal} {\bibinfo  {journal} {Am. J. Phys.}\ }\textbf {\bibinfo {volume}
  {50}},\ \bibinfo {pages} {66--71} (\bibinfo {year} {1982})}\BibitemShut
  {NoStop}%
\bibitem [{\citenamefont {McDermott}(1984)}]{mcdermott1984research}%
  \BibitemOpen
  \bibfield  {author} {\bibinfo {author} {\bibfnamefont {L.C.}\ \bibnamefont
  {McDermott}},\ }\bibfield  {title} {\enquote {\bibinfo {title} {Research on
  conceptual understanding in mechanics},}\ }\href@noop {} {\bibfield
  {journal} {\bibinfo  {journal} {Phys. Today}\ }\textbf {\bibinfo {volume}
  {37}},\ \bibinfo {pages} {24--32} (\bibinfo {year} {1984})}\BibitemShut
  {NoStop}%
\bibitem [{\citenamefont {Posner}\ \emph {et~al.}(1982)\citenamefont {Posner},
  \citenamefont {Strike}, \citenamefont {Hewson},\ and\ \citenamefont
  {Gertzog}}]{posner1982accommodation}%
  \BibitemOpen
  \bibfield  {author} {\bibinfo {author} {\bibfnamefont {G.J.}\ \bibnamefont
  {Posner}}, \bibinfo {author} {\bibfnamefont {K.A.}\ \bibnamefont {Strike}},
  \bibinfo {author} {\bibfnamefont {P.W.}\ \bibnamefont {Hewson}}, \ and\
  \bibinfo {author} {\bibfnamefont {W.A.}\ \bibnamefont {Gertzog}},\ }\bibfield
   {title} {\enquote {\bibinfo {title} {Accommodation of a scientific
  conception: {T}oward a theory of conceptual change},}\ }\href@noop {}
  {\bibfield  {journal} {\bibinfo  {journal} {Sci. Educ.}\ }\textbf {\bibinfo
  {volume} {66}},\ \bibinfo {pages} {211--227} (\bibinfo {year}
  {1982})}\BibitemShut {NoStop}%
\bibitem [{\citenamefont {Etkina}\ \emph {et~al.}(2005)\citenamefont {Etkina},
  \citenamefont {Mestre},\ and\ \citenamefont
  {O'Donnell}}]{etkina2005cognitive}%
  \BibitemOpen
  \bibfield  {author} {\bibinfo {author} {\bibfnamefont {E.}~\bibnamefont
  {Etkina}}, \bibinfo {author} {\bibfnamefont {J.}~\bibnamefont {Mestre}}, \
  and\ \bibinfo {author} {\bibfnamefont {A.}~\bibnamefont {O'Donnell}},\
  }\bibfield  {title} {\enquote {\bibinfo {title} {The cognitive revolution in
  educational psychology},}\ \ }(\bibinfo  {publisher} {Information Age
  Publishing},\ \bibinfo {address} {Greenwich, CT},\ \bibinfo {year}
  {2005})\ pp.\ \bibinfo {pages} {119--164}\BibitemShut {NoStop}%
\bibitem [{\citenamefont {Council}(2000)}]{national2000people}%
  \BibitemOpen
  \bibfield  {author} {\bibinfo {author} {\bibfnamefont {National~Research}\
  \bibnamefont {Council}},\ }\href@noop {} {\enquote {\bibinfo {title} {How
  {P}eople {L}earn: {B}rain, {M}ind, {Ex}perience, and {S}chool: {E}xpanded
  {E}dition},}\ }\bibinfo {howpublished} {The National Academies Press,
  Washington, DC} (\bibinfo {year} {2000})\BibitemShut {NoStop}%
\bibitem [{\citenamefont {Chi}\ and\ \citenamefont
  {Slotta}(1993)}]{chi1993ontological}%
  \BibitemOpen
  \bibfield  {author} {\bibinfo {author} {\bibfnamefont {M.T.H.}\ \bibnamefont
  {Chi}}\ and\ \bibinfo {author} {\bibfnamefont {J.D.}\ \bibnamefont
  {Slotta}},\ }\bibfield  {title} {\enquote {\bibinfo {title} {The ontological
  coherence of intuitive physics},}\ }\href@noop {} {\bibfield  {journal}
  {\bibinfo  {journal} {Cognition Instruct.}\ }\textbf {\bibinfo {volume}
  {10}},\ \bibinfo {pages} {249--260} (\bibinfo {year} {1993})}\BibitemShut
  {NoStop}%
\bibitem [{\citenamefont {Chi}\ \emph {et~al.}(1994)\citenamefont {Chi},
  \citenamefont {Slotta},\ and\ \citenamefont {De~Leeuw}}]{chi1994things}%
  \BibitemOpen
  \bibfield  {author} {\bibinfo {author} {\bibfnamefont {M.T.H.}\ \bibnamefont
  {Chi}}, \bibinfo {author} {\bibfnamefont {J.D.}\ \bibnamefont {Slotta}}, \
  and\ \bibinfo {author} {\bibfnamefont {N.}~\bibnamefont {De~Leeuw}},\
  }\bibfield  {title} {\enquote {\bibinfo {title} {From things to processes:
  {A} theory of conceptual change for learning science concepts},}\ }\href@noop
  {} {\bibfield  {journal} {\bibinfo  {journal} {Learn. Instruct.}\ }\textbf
  {\bibinfo {volume} {4}},\ \bibinfo {pages} {27--43} (\bibinfo {year}
  {1994})}\BibitemShut {NoStop}%
\bibitem [{\citenamefont {Slotta}\ \emph {et~al.}(1995)\citenamefont {Slotta},
  \citenamefont {Chi},\ and\ \citenamefont {Joram}}]{slotta1995assessing}%
  \BibitemOpen
  \bibfield  {author} {\bibinfo {author} {\bibfnamefont {J.D.}\ \bibnamefont
  {Slotta}}, \bibinfo {author} {\bibfnamefont {M.T.H}\ \bibnamefont {Chi}}, \
  and\ \bibinfo {author} {\bibfnamefont {E.}~\bibnamefont {Joram}},\ }\bibfield
   {title} {\enquote {\bibinfo {title} {Assessing students' misclassifications
  of physics concepts: {A}n ontological basis for conceptual change},}\
  }\href@noop {} {\bibfield  {journal} {\bibinfo  {journal} {Cognition
  Instruct.}\ }\textbf {\bibinfo {volume} {13}},\ \bibinfo {pages} {373--400}
  (\bibinfo {year} {1995})}\BibitemShut {NoStop}%
\bibitem [{\citenamefont {DiSessa}(1993)}]{disessa1993toward}%
  \BibitemOpen
  \bibfield  {author} {\bibinfo {author} {\bibfnamefont {A.A.}\ \bibnamefont
  {DiSessa}},\ }\bibfield  {title} {\enquote {\bibinfo {title} {Toward an
  epistemology of physics},}\ }\href@noop {} {\bibfield  {journal} {\bibinfo
  {journal} {Cognition Instruct.}\ }\textbf {\bibinfo {volume} {10}},\ \bibinfo
  {pages} {105--225} (\bibinfo {year} {1993})}\BibitemShut {NoStop}%
\bibitem [{\citenamefont {Disessa}\ and\ \citenamefont
  {Sherin}(1998)}]{disessa1998changes}%
  \BibitemOpen
  \bibfield  {author} {\bibinfo {author} {\bibfnamefont {A.A.}\ \bibnamefont
  {Disessa}}\ and\ \bibinfo {author} {\bibfnamefont {B.L.}\ \bibnamefont
  {Sherin}},\ }\bibfield  {title} {\enquote {\bibinfo {title} {What changes in
  conceptual change?}}\ }\href@noop {} {\bibfield  {journal} {\bibinfo
  {journal} {Int. J. Sci. Educ.}\ }\textbf {\bibinfo {volume} {20}},\ \bibinfo
  {pages} {1155--1191} (\bibinfo {year} {1998})}\BibitemShut {NoStop}%
\bibitem [{\citenamefont {Hammer}(1996)}]{hammer1996misconceptions}%
  \BibitemOpen
  \bibfield  {author} {\bibinfo {author} {\bibfnamefont {D.}~\bibnamefont
  {Hammer}},\ }\bibfield  {title} {\enquote {\bibinfo {title} {Misconceptions
  or p-prims: {H}ow may alternative perspectives of cognitive structure
  influence instructional perceptions and intentions},}\ }\href@noop {}
  {\bibfield  {journal} {\bibinfo  {journal} {J. Learn. Sci.}\ }\textbf
  {\bibinfo {volume} {5}},\ \bibinfo {pages} {97--127} (\bibinfo {year}
  {1996})}\BibitemShut {NoStop}%
\bibitem [{\citenamefont {diSessa}\ \emph {et~al.}(2004)\citenamefont
  {diSessa}, \citenamefont {Gillespie},\ and\ \citenamefont
  {Esterly}}]{gillespie2004coherence}%
  \BibitemOpen
  \bibfield  {author} {\bibinfo {author} {\bibfnamefont {A.A.}\ \bibnamefont
  {diSessa}}, \bibinfo {author} {\bibfnamefont {N.M.}\ \bibnamefont
  {Gillespie}}, \ and\ \bibinfo {author} {\bibfnamefont {J.B.}\ \bibnamefont
  {Esterly}},\ }\bibfield  {title} {\enquote {\bibinfo {title} {Coherence
  versus fragmentation in the development of the concept of force},}\
  }\href@noop {} {\bibfield  {journal} {\bibinfo  {journal} {Cognitive Sci.}\
  }\textbf {\bibinfo {volume} {28}},\ \bibinfo {pages} {843--900} (\bibinfo
  {year} {2004})}\BibitemShut {NoStop}%
\bibitem [{\citenamefont {Dufresne}\ \emph {et~al.}(2002)\citenamefont
  {Dufresne}, \citenamefont {Leonard},\ and\ \citenamefont
  {Gerace}}]{dufresne2002marking}%
  \BibitemOpen
  \bibfield  {author} {\bibinfo {author} {\bibfnamefont {R.J.}\ \bibnamefont
  {Dufresne}}, \bibinfo {author} {\bibfnamefont {W.J.}\ \bibnamefont
  {Leonard}}, \ and\ \bibinfo {author} {\bibfnamefont {W.J.}\ \bibnamefont
  {Gerace}},\ }\bibfield  {title} {\enquote {\bibinfo {title} {Making sense of
  students' answers to multiple-choice questions},}\ }\href@noop {} {\bibfield
  {journal} {\bibinfo  {journal} {Phys. Teach.}\ }\textbf {\bibinfo {volume}
  {40}},\ \bibinfo {pages} {174--180} (\bibinfo {year} {2002})}\BibitemShut
  {NoStop}%
\bibitem [{\citenamefont {Steinberg}\ and\ \citenamefont
  {Sabella}(1997)}]{steinberg1997performance}%
  \BibitemOpen
  \bibfield  {author} {\bibinfo {author} {\bibfnamefont {R.N.}\ \bibnamefont
  {Steinberg}}\ and\ \bibinfo {author} {\bibfnamefont {M.S.}\ \bibnamefont
  {Sabella}},\ }\bibfield  {title} {\enquote {\bibinfo {title} {Performance on
  multiple-choice diagnostics and complementary exam problems},}\ }\href@noop
  {} {\bibfield  {journal} {\bibinfo  {journal} {Phys. Teach.}\ }\textbf
  {\bibinfo {volume} {35}},\ \bibinfo {pages} {150--155} (\bibinfo {year}
  {1997})}\BibitemShut {NoStop}%
\bibitem [{\citenamefont {Newell}\ and\ \citenamefont
  {Simon}(1972)}]{simonnewell}%
  \BibitemOpen
  \bibfield  {author} {\bibinfo {author} {\bibfnamefont {A.}~\bibnamefont
  {Newell}}\ and\ \bibinfo {author} {\bibfnamefont {H.A.}\ \bibnamefont
  {Simon}},\ }\href@noop {} {\emph {\bibinfo {title} {Human Problem Solving}}}\
  (\bibinfo  {publisher} {Prentice-Hall},\ \bibinfo {address} {Englewood
  Cliffs, NJ},\ \bibinfo {year} {1972})\BibitemShut {NoStop}%
\bibitem [{\citenamefont {Ohlsson}(2012)}]{ohlsson2012problems}%
  \BibitemOpen
  \bibfield  {author} {\bibinfo {author} {\bibfnamefont {Stellan}\ \bibnamefont
  {Ohlsson}},\ }\bibfield  {title} {\enquote {\bibinfo {title} {The problems
  with problem solving: {R}eflections on the rise, current status, and possible
  future of a cognitive research paradigm},}\ }\href@noop {} {\bibfield
  {journal} {\bibinfo  {journal} {J. Prob. Solving}\ }\textbf {\bibinfo
  {volume} {5}},\ \bibinfo {pages} {7} (\bibinfo {year} {2012})}\BibitemShut
  {NoStop}%
\bibitem [{\citenamefont {Larkin}\ \emph
  {et~al.}(1980{\natexlab{a}})\citenamefont {Larkin}, \citenamefont
  {McDermott}, \citenamefont {Simon},\ and\ \citenamefont
  {Simon}}]{larkin1980expert}%
  \BibitemOpen
  \bibfield  {author} {\bibinfo {author} {\bibfnamefont {J.}~\bibnamefont
  {Larkin}}, \bibinfo {author} {\bibfnamefont {J.}~\bibnamefont {McDermott}},
  \bibinfo {author} {\bibfnamefont {D.P.}\ \bibnamefont {Simon}}, \ and\
  \bibinfo {author} {\bibfnamefont {H.A.}\ \bibnamefont {Simon}},\ }\bibfield
  {title} {\enquote {\bibinfo {title} {Expert and novice performance in solving
  physics problems},}\ }\href@noop {} {\bibfield  {journal} {\bibinfo
  {journal} {Science}\ }\textbf {\bibinfo {volume} {208}},\ \bibinfo {pages}
  {1335--1342} (\bibinfo {year} {1980}{\natexlab{a}})}\BibitemShut {NoStop}%
\bibitem [{\citenamefont {Larkin}\ \emph
  {et~al.}(1980{\natexlab{b}})\citenamefont {Larkin}, \citenamefont
  {McDermott}, \citenamefont {Simon},\ and\ \citenamefont
  {Simon}}]{larkin1980models}%
  \BibitemOpen
  \bibfield  {author} {\bibinfo {author} {\bibfnamefont {J.H.}\ \bibnamefont
  {Larkin}}, \bibinfo {author} {\bibfnamefont {J.}~\bibnamefont {McDermott}},
  \bibinfo {author} {\bibfnamefont {D.P.}\ \bibnamefont {Simon}}, \ and\
  \bibinfo {author} {\bibfnamefont {H.A.}\ \bibnamefont {Simon}},\ }\bibfield
  {title} {\enquote {\bibinfo {title} {Models of competence in solving physics
  problems},}\ }\href@noop {} {\bibfield  {journal} {\bibinfo  {journal}
  {Cognitive Sci.}\ }\textbf {\bibinfo {volume} {4}},\ \bibinfo {pages}
  {317--345} (\bibinfo {year} {1980}{\natexlab{b}})}\BibitemShut {NoStop}%
\bibitem [{\citenamefont {Madsen}\ \emph {et~al.}(2013)\citenamefont {Madsen},
  \citenamefont {McKagan},\ and\ \citenamefont {Sayre}}]{madsen_gender_2013}%
  \BibitemOpen
  \bibfield  {author} {\bibinfo {author} {\bibfnamefont {A.}~\bibnamefont
  {Madsen}}, \bibinfo {author} {\bibfnamefont {S.B.}\ \bibnamefont {McKagan}},
  \ and\ \bibinfo {author} {\bibfnamefont {E.}~\bibnamefont {Sayre}},\
  }\bibfield  {title} {\enquote {\bibinfo {title} {Gender gap on concept
  inventories in physics: {What} is consistent, what is inconsistent, and what
  factors influence the gap?}}\ }\href {\doibase 10.1103/PhysRevSTPER.9.020121}
  {\bibfield  {journal} {\bibinfo  {journal} {Phys. Rev. Phys. Educ. Res.}\
  }\textbf {\bibinfo {volume} {9}},\ \bibinfo {pages} {020121} (\bibinfo {year}
  {2013})}\BibitemShut {NoStop}%
\bibitem [{\citenamefont {Henderson}\ \emph {et~al.}(2017)\citenamefont
  {Henderson}, \citenamefont {Stewart}, \citenamefont {Stewart}, \citenamefont
  {Michaluk},\ and\ \citenamefont {Traxler}}]{henderson2017}%
  \BibitemOpen
  \bibfield  {author} {\bibinfo {author} {\bibfnamefont {R.}~\bibnamefont
  {Henderson}}, \bibinfo {author} {\bibfnamefont {G.}~\bibnamefont {Stewart}},
  \bibinfo {author} {\bibfnamefont {J.}~\bibnamefont {Stewart}}, \bibinfo
  {author} {\bibfnamefont {L.}~\bibnamefont {Michaluk}}, \ and\ \bibinfo
  {author} {\bibfnamefont {A.}~\bibnamefont {Traxler}},\ }\bibfield  {title}
  {\enquote {\bibinfo {title} {Exploring the gender gap in the {C}onceptual
  {S}urvey of {E}lectricity and {M}agnetism},}\ }\href {\doibase
  10.1103/PhysRevPhysEducRes.13.020114} {\bibfield  {journal} {\bibinfo
  {journal} {Phys. Rev. Phys. Educ. Res.}\ }\textbf {\bibinfo {volume} {13}},\
  \bibinfo {pages} {020114} (\bibinfo {year} {2017})}\BibitemShut {NoStop}%
\bibitem [{\citenamefont {Traxler}\ \emph {et~al.}(2016)\citenamefont
  {Traxler}, \citenamefont {Cid}, \citenamefont {Blue},\ and\ \citenamefont
  {Barthelemy}}]{traxler_enriching_2016}%
  \BibitemOpen
  \bibfield  {author} {\bibinfo {author} {\bibfnamefont {A.L.}\ \bibnamefont
  {Traxler}}, \bibinfo {author} {\bibfnamefont {X.C.}\ \bibnamefont {Cid}},
  \bibinfo {author} {\bibfnamefont {J.}~\bibnamefont {Blue}}, \ and\ \bibinfo
  {author} {\bibfnamefont {R.}~\bibnamefont {Barthelemy}},\ }\bibfield  {title}
  {\enquote {\bibinfo {title} {Enriching gender in physics education research:
  {A} binary past and a complex future},}\ }\href {\doibase
  10.1103/PhysRevPhysEducRes.12.020114} {\bibfield  {journal} {\bibinfo
  {journal} {Phys. Rev. Phys. Educ. Res.}\ }\textbf {\bibinfo {volume} {12}},\
  \bibinfo {pages} {020114} (\bibinfo {year} {2016})}\BibitemShut {NoStop}%
\bibitem [{\citenamefont {Mazur}(1997)}]{mazur1996peer}%
  \BibitemOpen
  \bibfield  {author} {\bibinfo {author} {\bibfnamefont {E.}~\bibnamefont
  {Mazur}},\ }\href@noop {} {\emph {\bibinfo {title} {Peer {Instruction}: {A}
  {U}ser's {M}anual}}}\ (\bibinfo  {publisher} {Prentice Hall},\ \bibinfo
  {address} {Upper Saddle River, NJ},\ \bibinfo {year} {1997})\BibitemShut
  {NoStop}%
\bibitem [{phy()}]{physport}%
  \BibitemOpen
  \href {https://www.physport.org/} {\enquote {\bibinfo {title} {Physport},}\
  }\bibinfo {note} {\url{https://www.physport.org}. Accessed
  8/8/2017.}\BibitemShut {Stop}%
\bibitem [{usn()}]{usnews}%
  \BibitemOpen
  \href {https://premium.usnews.com/best-colleges} {\enquote {\bibinfo {title}
  {{US N}ews \& {W}orld {R}eport: {E}ducation},}\ }\bibinfo {howpublished} {US
  News and World Report, Washington, DC},\ \bibinfo {note}
  {\url{https://premium.usnews.com/best-colleges}. Accessed
  4/30/2017.}\BibitemShut {Stop}%
\bibitem [{\citenamefont {van~der Linden}(2016)}]{crcbook}%
  \BibitemOpen
  \bibfield  {author} {\bibinfo {author} {\bibfnamefont {W.J.}\ \bibnamefont
  {van~der Linden}},\ }\bibfield  {title} {\enquote {\bibinfo {title}
  {{U}nidimensional {L}ogistic {R}esponse {M}odels},}\ }in\ \href@noop {}
  {\emph {\bibinfo {booktitle} {{H}andbook of {I}tem {R}esponse {T}heory}}},\
  Vol.~\bibinfo {volume} {1}\ (\bibinfo  {publisher} {CRC Press, Taylor \&
  Francis Group},\ \bibinfo {address} {New York, NY},\ \bibinfo {year} {2016})\
  pp.\ \bibinfo {pages} {13--30}\BibitemShut {NoStop}%
\bibitem [{\citenamefont {Hu}\ and\ \citenamefont {Bentler}(1999)}]{hu1999}%
  \BibitemOpen
  \bibfield  {author} {\bibinfo {author} {\bibfnamefont {L.}~\bibnamefont
  {Hu}}\ and\ \bibinfo {author} {\bibfnamefont {P.M.}\ \bibnamefont
  {Bentler}},\ }\bibfield  {title} {\enquote {\bibinfo {title} {Cutoff criteria
  for fit indexes in covariance structure analysis: Conventional criteria
  versus new alternatives},}\ }\href@noop {} {\bibfield  {journal} {\bibinfo
  {journal} {Struct. Equ. Modeling}\ }\textbf {\bibinfo {volume} {6}},\
  \bibinfo {pages} {1--55} (\bibinfo {year} {1999})}\BibitemShut {NoStop}%
\bibitem [{sup()}]{supp}%
  \BibitemOpen
  \href@noop {} {}\bibinfo {note} {See Supplemental Material at [URL will be
  inserted by publisher] for traditional factor analysis, 3 and 5 factor MIRT
  factor models, and the constrained MIRT model with the factor loading on all
  items.}\BibitemShut {Stop}%
\bibitem [{\citenamefont {{R Core Team}}(2013)}]{rsoftware}%
  \BibitemOpen
  \bibfield  {author} {\bibinfo {author} {\bibnamefont {{R Core Team}}},\
  }\href {http://www.R-project.org/} {\emph {\bibinfo {title} {R: A Language
  and Environment for Statistical Computing}}},\ \bibinfo {organization} {R
  Foundation for Statistical Computing},\ \bibinfo {address} {Vienna, Austria}
  (\bibinfo {year} {2013})\BibitemShut {NoStop}%
\bibitem [{\citenamefont {Chalmers}(2012)}]{mirt}%
  \BibitemOpen
  \bibfield  {author} {\bibinfo {author} {\bibfnamefont {R.P.}\ \bibnamefont
  {Chalmers}},\ }\bibfield  {title} {\enquote {\bibinfo {title} {mirt: A
  multidimensional item response theory package for the {R} environment},}\
  }\href@noop {} {\bibfield  {journal} {\bibinfo  {journal} {J. Stat. Soft.}\
  }\textbf {\bibinfo {volume} {48}},\ \bibinfo {pages} {1--29} (\bibinfo {year}
  {2012})}\BibitemShut {NoStop}%
\bibitem [{\citenamefont {Epskamp}\ \emph {et~al.}(2012)\citenamefont
  {Epskamp}, \citenamefont {Cramer}, \citenamefont {Waldorp}, \citenamefont
  {Schmittmann},\ and\ \citenamefont {Borsboom}}]{qgraph}%
  \BibitemOpen
  \bibfield  {author} {\bibinfo {author} {\bibfnamefont {S.}~\bibnamefont
  {Epskamp}}, \bibinfo {author} {\bibfnamefont {A.O.J.}\ \bibnamefont
  {Cramer}}, \bibinfo {author} {\bibfnamefont {J.L.}\ \bibnamefont {Waldorp}},
  \bibinfo {author} {\bibfnamefont {V.D.}\ \bibnamefont {Schmittmann}}, \ and\
  \bibinfo {author} {\bibfnamefont {D.}~\bibnamefont {Borsboom}},\ }\bibfield
  {title} {\enquote {\bibinfo {title} {{qgraph}: Network visualizations of
  relationships in psychometric data},}\ }\href
  {http://www.jstatsoft.org/v48/i04/} {\bibfield  {journal} {\bibinfo
  {journal} {J. Stat. Soft.}\ }\textbf {\bibinfo {volume} {48}},\ \bibinfo
  {pages} {1--18} (\bibinfo {year} {2012})}\BibitemShut {NoStop}%
\bibitem [{\citenamefont {Canty}\ and\ \citenamefont {Ripley}(2017)}]{boot1}%
  \BibitemOpen
  \bibfield  {author} {\bibinfo {author} {\bibfnamefont {A.}~\bibnamefont
  {Canty}}\ and\ \bibinfo {author} {\bibfnamefont {B.D.}\ \bibnamefont
  {Ripley}},\ }\href@noop {} {\emph {\bibinfo {title} {boot: Bootstrap {R}
  {(S-Plus)} Functions}}} (\bibinfo {year} {2017}),\ \bibinfo {note} {{R}
  package version 1.3-20}\BibitemShut {NoStop}%
\bibitem [{\citenamefont {Davison}\ and\ \citenamefont
  {Hinkley}(1997)}]{boot2}%
  \BibitemOpen
  \bibfield  {author} {\bibinfo {author} {\bibfnamefont {A.C.}\ \bibnamefont
  {Davison}}\ and\ \bibinfo {author} {\bibfnamefont {D.V.}\ \bibnamefont
  {Hinkley}},\ }\href {http://statwww.epfl.ch/davison/BMA/} {\emph {\bibinfo
  {title} {Bootstrap Methods and Their Applications}}}\ (\bibinfo  {publisher}
  {Cambridge University Press},\ \bibinfo {address} {Cambridge, UK},\ \bibinfo
  {year} {1997})\BibitemShut {NoStop}%
\bibitem [{\citenamefont {Cronbach}\ and\ \citenamefont
  {Meehl}(1955)}]{cronbach1955construct}%
  \BibitemOpen
  \bibfield  {author} {\bibinfo {author} {\bibfnamefont {L.J.}\ \bibnamefont
  {Cronbach}}\ and\ \bibinfo {author} {\bibfnamefont {P.E.}\ \bibnamefont
  {Meehl}},\ }\bibfield  {title} {\enquote {\bibinfo {title} {Construct
  validity in psychological tests.}}\ }\href@noop {} {\bibfield  {journal}
  {\bibinfo  {journal} {Psychol. Bull.}\ }\textbf {\bibinfo {volume} {52}},\
  \bibinfo {pages} {281} (\bibinfo {year} {1955})}\BibitemShut {NoStop}%
\bibitem [{\citenamefont {Clark}\ and\ \citenamefont
  {Watson}(1995)}]{clark1995constructing}%
  \BibitemOpen
  \bibfield  {author} {\bibinfo {author} {\bibfnamefont {L.A.}\ \bibnamefont
  {Clark}}\ and\ \bibinfo {author} {\bibfnamefont {D.}~\bibnamefont {Watson}},\
  }\bibfield  {title} {\enquote {\bibinfo {title} {Constructing validity:
  {B}asic issues in objective scale development.}}\ }\href@noop {} {\bibfield
  {journal} {\bibinfo  {journal} {Psychol. Assessment}\ }\textbf {\bibinfo
  {volume} {7}},\ \bibinfo {pages} {309} (\bibinfo {year} {1995})}\BibitemShut
  {NoStop}%
\end{thebibliography}

%

\end{document}